\documentclass[aps,pra,superscriptaddress,stfloats,twocolumn,10pt]{revtex4-1}

\usepackage[latin1]{inputenc}

\usepackage{amssymb,amsmath}
\usepackage{graphicx}
\usepackage{placeins}
\usepackage{epstopdf}
\usepackage{subfigure}
\usepackage[none]{hyphenat}
\usepackage{hyperref}
\usepackage{natbib}
\usepackage{color}
\usepackage[normalem]{ulem}

\definecolor{darkred}{rgb}{0.45,0.,0.0}
\definecolor{darkblue}{rgb}{0.,0.,0.75}

\hypersetup{backref,colorlinks=true,linkcolor=blue,citecolor=blue,urlcolor=blue,linktocpage=true,breaklinks}
\usepackage{cancel}
\usepackage{float}
\DeclareMathAlphabet{\mathbfit}{OML}{cmm}{b}{it}

\renewcommand{\vec}[1]{\mathbf{#1}}

\definecolor{dgreen}{rgb}{0.0, 0.5, 0.0}

\begin{document}
\title{Rotation of cold molecular ions inside a Bose-Einstein condensate}

\author{Bikashkali Midya}
\email{bikashkali.midya@ist.ac.at}
\affiliation{IST Austria  (Institute of Science and Technology Austria), Am Campus 1, 3400 Klosterneuburg, Austria}

\author{ Micha{\l}  Tomza}
\email{michal.tomza@icfo.es}
\affiliation{ICFO-Institut de Ci\`{e}ncies Fot\`{o}niques, The Barcelona Institute of Science and Technology, Barcelona, Spain}

\author{Richard Schmidt}
\email{richard.schmidt@cfa.harvard.edu}
 \affiliation{ITAMP, Havard-Smithsonian Center for Astrophysics, Cambridge, MA 02138, USA }
   \affiliation{Physics Department, Harvard University, 17 Oxford Street, Cambridge, MA 02138, USA} 

\author{Mikhail Lemeshko}
\email{mikhail.lemeshko@ist.ac.at}
\affiliation{IST Austria  (Institute of Science and Technology Austria), Am Campus 1, 3400 Klosterneuburg, Austria}

\begin{abstract}
 We use recently developed angulon theory [Phys.~Rev.~Lett.~\textbf{114}, 203001~(2015)] to study the rotational spectrum of a cyanide molecular anion immersed into Bose-Einstein condensates of rubidium and strontium. Based on \textit{ab initio} potential energy surfaces, we provide a detailed study of the rotational Lamb shift and many-body-induced fine structure which arise due to dressing of molecular rotation by a field of phonon excitations. We demonstrate that the magnitude of these effects is large enough in order to be observed in modern experiments on cold molecular ions. Furthermore, we introduce a novel method to construct pseudopotentials starting from the \textit {ab initio} potential energy surfaces, which provides a means to obtain effective coupling constants for low-energy polaron models.
\end{abstract}


 \maketitle

{\it Introduction.} Recently, cold molecular ions came about as a versatile platform to study single-, few-, and many-particle quantum processes~\cite{WillitschIRPC12,WillitschPCCP2008,LemKreDoyKais13}.  As opposed to neutral molecules~\cite{LemKreDoyKais13, *KreStwFrieColdMol}, the degree of freedom used to manipulate molecular ions -- their charge -- is  effectively decoupled from their internal structure, and therefore even complex species can be trapped and cooled down to millikelvin translational temperatures~\cite{WillitschIRPC12}. Molecular  ions can be prepared in a selected rovibrational state by optical pumping~\cite{StaanumNatPhys10,Schneider2010,Lien2014} or sympathetic cooling of state-selected ions~\cite{TongPRL10}; they can be trapped for a few hours, with the rotational state lifetimes exceeding 15~min~\cite{TongPRL10}.  
 
The interest to the topic is driven, however, by numerous potential applications, such as quantum information processing~\cite{Petit2013,ShiNJP13}, astrochemistry~\cite{LarssonRPP12, FortenberryJPCA15}, as well as  taming reactive collisions in the quantum regime~\cite{HallPRL12,HojbjerrePRA08}.  Furthermore, trapped and cold molecular ions play a crucial role in precision spectroscopy~\cite{BresselPRL12,GerToWillNatPhys2014}, as well as tests of fundamental physics, ranging from parity violation effects~\cite{BorschevskyPRA2012} to search for an  electric dipole moment of the electron~\cite{LeanhardtJmolSpec11, LohScience13}, and possible time variation of the electron-to-proton mass ratio~\cite{WolfNature2016, BiesheuvelNatComm16}.

Last but not least, ensembles of interacting ultracold polar molecules and ions are promising candidates to simulate complex many-body states of matter. Such simulations hold a potential to uncover the elusive nature of strongly-correlated quantum systems, such as high-temperature superconductors~\cite{LewensteinBook12}. Recently, there appeared numerous theoretical proposals on simulating many-particle physics with ultracold molecules~\cite{LewensteinBook12,Carr09,LemKreDoyKais13,JinYeCRev12,KreStwFrieColdMol}, some of which have already been realized in laboratory~\cite{YanNature13}. Many of such proposals make use of long-range dipole-dipole interactions between particles in a high-density molecular sample. For neutral molecules, the latter is sometimes challenging to engineer experimentally, partially because of the reactive collisions between the species~\cite{YeARPC2014,KreStwFrieColdMol, LemKreDoyKais13}.  For molecular ions, on the other hand, strong dipole-dipole interactions are precluded by large distances between the species in an ion trap~\cite{WillitschIRPC12}.
 
 The above mentioned problem, however, does not arise if one deals with a single molecule (or a very dilute molecular sample) coupled to an atomic Bose-Einstein condensate (BEC)~\cite{RellergertNat13, StoecklinNatComm16, HauserNatPhys15,OttoPRL08,DeiglmayrPRA12}. Recently, it was shown that in such conditions, molecular rotational motion becomes dressed by a field of many-body excitations, giving rise to quasiparticles of a new kind -- the `angulons'~\cite{SchmidtLem15, SchmidtLem16, LemSchmidtChapter}. In a way, angulons represent a rotational analogue of polarons~\cite{LandauPolaron,FrohlichAdvPhys54,FeynmanPR55,Devreese15,GrusdtCourse15}, actively studied experimentally in ultracold quantum gases~\cite{ChikkaturPRL00, SchirotzekPRL09, PalzerPRL09, KohstallNature12, KoschorreckNature12, SpethmannPRL12, FukuharaNatPhys13, ScellePRL13, Cetina15, MassignanRPP14,Jorgensen2016, Hu16, Cetina2016}. However, the non-Abelian algebra and discrete energy spectrum associated with quantum rotations makes the angulon physics drastically different compared to any polaron problem, where  the translational motion of an impurity is studied. 

 Here we undertake a detailed quantitative study of  the effects predicted in Refs.~\cite{SchmidtLem15, SchmidtLem16} and show that the latter are within reach in current experiments on cold molecular ions. As a concrete experimental system we consider a single cyanide (CN$^-$) anion immersed into an atomic BEC of $^{87}$Rb and $^{84}$Sr. We illustrate our technique with molecular anions, since those have been used in the context of hybrid ion-atom setups in order to achieve sympathetic cooling~\cite{DeiglmayrPRA12}. Furthermore, negative ions have been used to study cold rotationally inelastic~\cite{HauserNatPhys15,GonzalezCP15, GonzalezNJP15} and reactive~\cite{OttoPRL08, MikoschIRPC, WesterHandSpec11} collisions, and are highly relevant for astrochemistry~\cite{LarssonRPP12, FortenberryJPCA15}. It is worth noting, however, that the predicted effects are not expected to change substantially, either  qualitatively or quantitatively, for the case of positive molecular ions.

{\it The angulon quasiparticle.}  We  start  by outlining the basics of the angulon theory which describes a rotating quantum impurity immersed into  a weakly-interacting BEC. For simplicity, we consider a linear-rotor molecular ion, however, the theory can be generalized to more complex species. The microscopic Hamiltonian for such a system was derived in Ref.~\cite{SchmidtLem15} and is given by
\begin{equation}\label{hamiltonian}
\begin{split}
 \widehat{H} &= B \hat{\vec{J}}^2 + \sum\limits_{k\lambda\mu} \omega_k \hat{b}^\dag_{k\lambda\mu} \hat{b}_{k\lambda\mu} \\ 
 &+ \sum\limits_{k\lambda\mu} U_\lambda(k) ~ [Y^*_{\lambda\mu}(\hat{\theta},\hat{\phi})\hat{b}^\dag_{k\lambda\mu}+Y_{\lambda\mu}(\hat{\theta},\hat{\phi}) \hat{b}_{k\lambda\mu}],
 \end{split}
\end{equation}
where $\sum_k \equiv \int dk$ and $\hbar \equiv 1$. The first term of the Hamiltonian \eqref{hamiltonian} corresponds to the rotational kinetic
energy of the linear-rotor, with $B$ the rotational constant and $\hat{\vec{J}}$ the angular momentum operator. Since we focus on cold molecular ions, we neglect the translational kinetic energy of the impurity. The bare 
impurity eigenstates (in the absence of an external field) are characterised by the $(2j+1)$-fold degenerate levels $ | j,m\rangle$ with energies $E_j = B j(j + 1)$, where $j$ is the angular momentum quantum number, and $m$ its projection onto the laboratory $z$-axis. The second term of Eq. \eqref{hamiltonian} represents the kinetic energy
of the bosonic bath with Bogoliubov dispersion relation, $\omega_k = \sqrt{\varepsilon_k  (\varepsilon_k + 2 n g_{bb})}$, where  $\varepsilon_k = k^2/2m_b$ is the kinetic energy of a free boson of mass $m_b$  and $n$ is the particle number density in the BEC. Here $g_{bb}=4 \pi a_{bb}/m_b$ represents the strength of the boson-boson contact interactions and is defined in terms of the scattering length, $a_{bb}$, between them. The operators $\hat{b}^\dag_{k\lambda\mu}$ and $\hat{b}_{k\lambda\mu}$ are the phonon creation and annihilation operators, respectively. They are defined in the angular momentum basis as given by three quantum numbers: the magnitude of the linear momentum $k=|\vec{k}|$, the angular momentum, $\lambda$, and its projection, $\mu$, onto the $z$-axis, see Refs.~\cite{SchmidtLem15, SchmidtLem16, LemSchmidtChapter} for details.

The last term of the Hamiltonian describes the interaction between the impurity and the bath, where the angular momentum dependent interaction potential in Fourier space is given by:
\begin{equation}\label{potential_FT}
U_\lambda(k) =  \sqrt{\frac{32 \pi n k^2 {\varepsilon}_k}{{\omega}_k (2\lambda+1)^2}} {\int_{ 0}}^\infty dr ~ r^2 f_\lambda(r) j_\lambda(kr).
\end{equation}
Here $j_{\lambda}(k r)$ is the spherical Bessel function. The potential $U_\lambda(k)$ depends on the microscopic details of the two-body interaction between the impurity and the bosons through the functions $f_\lambda(r)$. In the molecular frame, the axially-symmetric molecule-boson interaction potential can be expanded in terms of Legendre polynomials $P_\lambda(\cos \theta')$ as follows
\begin{equation}\label{eq:Vn}
V_{\mathrm{imp-bos}} (\vec{r}')  = \sum_{\lambda} f_{\lambda}(r') P_{\lambda}(\cos\theta'),
\end{equation}
where the $f_\lambda (r')$ represents the shape of the potential in the respective angular momentum channel $\lambda$, $r'$ is the distance between
the atom and the centre-of-mass of the molecular ion, and $\theta'$ is the angle between the molecular axis and the axis connecting the atom with the centre-of-mass of the molecular ion. This potential \eqref{eq:Vn} can  be readily transformed to the laboratory frame by making use of the Wigner rotation matrices, which results in the third term of Eq. \eqref{hamiltonian}~\cite{SchmidtLem15, SchmidtLem16, LemSchmidtChapter}.

In order to reveal the rotational energies of the combined molecule-bath system, we use the following variational ansatz for the many-body quantum state
\begin{align}\label{ansatz}
|\psi\rangle =  Z^{1/2}_{LM} |0\rangle |LM\rangle + \underset{jm}{\sum\limits_{k\lambda\mu}} \beta_{k\lambda j} C^{LM}_{jm,\lambda \mu} ~ \hat{b}^\dag_{k\lambda\mu} |0\rangle |jm\rangle,
\end{align}
where $| 0\rangle$ is the BEC vacuum state, and $Z^{1/2}_{LM}$  and $\beta_{k\lambda j}$ are the variational parameters. The total angular momentum, $L$, and its projection, $M$, represent the conserved quantities of the problem, and this concervation law is built into the variational ansatz by the  Clebsch-Gordan coefficient, $C^{LM}_{jm,\lambda \mu}$. In the absence of external fields, the observed phenomena are independent of the $M$ quantum number, which we will omit hereafter.
For each value of $L$, the variational energies, $E_L$, are obtained by minimisation  of $E = \langle \psi | H | \psi \rangle / \langle \psi | \psi \rangle$ subject to the normalisation constraint, $Z_{LM} + \sum_{k\lambda j} |\beta_{k\lambda j}|^2 =1$. Furthermore, in Refs.~\cite{SchmidtLem15, LemSchmidtChapter} it has been shown that the variational approach to the angulon problem is equivalent to a diagrammatic one, as given by the following Dyson  equation for the angulon Green's function, $G_L(E)$: 
\begin{align}
[G_L(E)]^{-1} = [G_L^0(E)]^{-1} - \Sigma_L(E) = 0.
\end{align}
Here $[G_L^0(E)]^{-1} = B L (L+1) - E$ is the inverse Green's function of a free molecular ion and 
\begin{align}
\Sigma_L(E) = \sum\limits_{k\lambda j} \frac{2\lambda +1}{4 \pi} \frac{U_\lambda(k)^2 \left[C_{L0,\lambda0}^{j0}\right]^2}{B j(j+1) - E + \omega_k}
\end{align}
is the self energy, which arises due to the interaction between the molecular ion and the BEC. As a result, in addition to the ground-state properties furnished by any variational approach, this technique allows to  determine the angulon spectral function: 
\begin{equation}\label{eq:AE}
A_L(E) = \mathrm{Im} \left[G_L(E + i 0^+ )\right].
\end{equation}
The angulon spectral function determines not only the quasiparticle properties of the angulon but also allows to extract its entire excitation spectrum and represent the central object of this study.

{\it Anion-atom two-body interaction potentials.} In this section we provide details on the \textit{ab initio} calculations of the potential energy surfaces used to evaluate the components $f_{\lambda}(r')$ in Eq.~\eqref{eq:Vn}. The interaction between the CN$^-$ molecular ion in the  $^1\Sigma^+$ electronic ground state and the Rb atom in the $^2S$ ground state (the Sr atom in the $^1S$ state) results in one electronic state of  $^2A'$ ($^1A'$) symmetry; furthermore, considered systems are stable against any two-body chemical losses. Within the rigid rotor approximation, we fix the internuclear distance of CN$^-$ at the experimental value of $2.224\,$a.u.~\cite{BradforthJCP93} and define the $z'$ axis pointing from N to C. We calculate the potential-energy surface for the CN$^-$+Rb (CN$^-$+Sr) with the spin-restricted open-shell (closed-shell) coupled-cluster method restricted to single, double, and noniterative triple excitations [CCSD(T)]~\cite{MusialRMP07}, starting from the restricted open-shell (closed-shell) Hartree-Fock orbitals. The interaction energies are obtained with the supermolecule method with
the basis set superposition error corrected by using the counterpoise correction~\cite{BoysMP70}.

The N and C atoms are described using the augmented correlation-consistent polarized valence quintuple-$\zeta$ quality basis sets (aug-cc-pV5Z)~\cite{DunningJCP89}.
The scalar relativistic effects in Rb and Sr atoms are included by employing the small-core relativistic energy-consistent pseudopotentials (ECP) to replace the inner-shells electrons~\cite{DolgCR12}. The Rb and Sr atoms are described using the ECP28MDF pseudopotentials~\cite{LimJCP05,LimJCP06} and the $[14s14p7d6f1g]$~\cite{TomzaMP13} and $[14s11p6d5f4g]$~\cite{TomzaPCCP11} basis sets, respectively, obtained by augmenting the basis sets suggested in Refs.~\cite{LimJCP05,LimJCP06}. 
All electronic structure calculations are performed with the \textsc{Molpro} package of \textit{ab initio} programs \cite{Molpro}.  

The potential energy surfaces, $V_{\mathrm{imp-bos}}(r',\theta')$, are calculated on a two-dimensional grid consisting of 25 points in the ion-atom distance $r'$ with values between 3.5$\,$a.u.\ and 25$\,$a.u., and 12 points in the angle $\theta'$ with values between $0^\circ$ and $180^\circ$ chosen to be the quadratures for the Legendre polynomial of order $12$. Next, the calculated surfaces are expanded into Legendre polynomials as given by Eq.~\eqref{eq:Vn} and the Legendre components $f_\lambda(r')$ are obtained by integrating out the \textit{ab initio} points.  For CN$^-$+Rb (CN$^-$+Sr), the global minimum of the obtained potential energy surface has a depth of $D_e=0.0441$~a.u.\ and is  located at $r_e'=5.44\,$a.u., $\theta_e'=108^\circ$ ($D_e=0.0635$~a.u.\ at $r_e'=5.67\,$a.u., $\theta_e'=180^\circ$) and  for the secondary minimum $D_e=0.0438$~a.u.\ at $r_e'=6.14\,$a.u., $\theta_e'=180^\circ$ ($D_e=0.0575$~a.u.\ at $r_e'=6.20\,$a.u., $\theta_e'=0^\circ$).

\begin{figure}[t]
 \includegraphics[width=1\columnwidth,height=3.75cm]{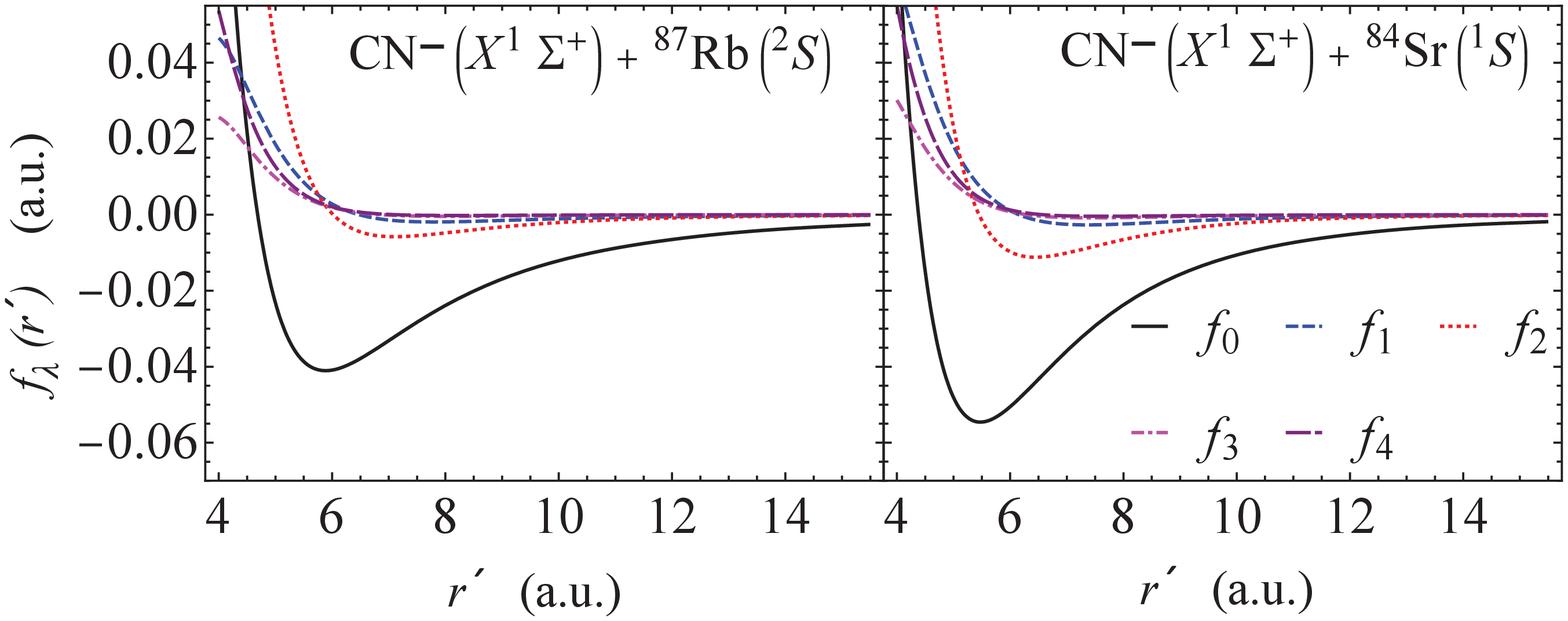} 
 \caption{(Color online) First five Legendre components, $f_\lambda(r')$, of the ground-state potential energy surface for CN$^-$($X^1\Sigma^+$) + Rb($^2S$) and CN$^-$($X^1\Sigma^+$) + Sr($^1S$).}\label{fig:potential}
\end{figure}
\FloatBarrier

\begingroup
\squeezetable
\begin{table}[t]
\caption{Induction and dispersion coefficients (in a.u.) describing the long-range part of the CN$^- +$Rb and CN$^- +$Sr potentials.\label{tab:Cn}} 
\begin{ruledtabular}
\begin{tabular}{lrrrrrr}
System &  $C^{\mathrm{ind}}_4\,$ & $C^{\mathrm{ind}}_5\,$ & $C^{\mathrm{ind}}_6\,$ &  $\Delta C^{\mathrm{ind}}_6\,$ &  $C^{\mathrm{disp}}_6\,$ & $\Delta C^{\mathrm{disp}}_6\,$\\
\hline
CN$^-$+Rb &     159.8 &  165 & 3310 & 3535 &  817 &  80.7 \\
CN$^-$+Sr &      99.6 &  103 & 2289 & 2202 & 767 &  75.0 \\
\end{tabular}
\label{tab:spec}
\end{ruledtabular}
\end{table}
\endgroup

In the ultracold regime, a dominant role in the description of the scattering between impurity and atoms is played by the interaction potential at large intermolecular distances, which can be represented as~\cite{JeziorskiCR94}: 
\begin{equation}\label{eq:En}
\begin{split}
V_{\mathrm{imp-bos}}(r' \to \infty,\theta') \approx -\frac{C^\textrm{ind}_4}{r'^4}-\frac{C^\textrm{ind}_5}{r'^5}\cos\theta'-\frac{C^\textrm{ind}_6}{r'^6}\\ -\frac{C^\textrm{disp}_6}{r'^6}
-\left(\frac{\Delta C^\textrm{ind}_6}{r'^6}+\frac{\Delta C^\textrm{disp}_6}{r'^6}\right)P_2(\cos\theta')+\dots\,,
\end{split}
\end{equation}   
where the leading long-range induction coefficient, $C^{\mathrm{ind}}_4=\frac{1}{2}q^2\alpha_\textrm{atom}$, is given by the static electric dipole polarizability of the atom, $\alpha_\textrm{atom}$, and the charge of the ion, $q$. The following induction coefficients $C^{\mathrm{ind}}_5=2q\alpha_\textrm{atom}d_\textrm{ion}$, $C^{\mathrm{ind}}_6=\frac{1}{2}q^2\beta_\textrm{atom}+\alpha_\textrm{atom}d^2_\textrm{ion}$, and
$\Delta C^{\mathrm{ind}}_6=2q\alpha_\textrm{atom} \Theta_\textrm{ion}+\alpha_\textrm{atom}d^2_\textrm{ion}$, are also expressed through the electric properties of monomers: the permanent electric dipole moment of the ion, $d_\textrm{ion}$, the permanent quadrupole moment of the ion, $\Theta_\textrm{ion}$,  as well as the static electric quadrupole polarizability of the atom, $\beta_\textrm{atom}$.
The leading long-range dispersion coefficients are given by 
$C^\textrm{disp}_6=\frac{3}{\pi}\int_0^\infty  \bar{\alpha}_{\textrm{ion}}(i\omega)\bar{\alpha}_\textrm{atom}(i\omega)d\omega$,
$\Delta C^\textrm{disp}_6=\frac{1}{\pi}\int_0^\infty \Delta\alpha_{\textrm{ion}}(i\omega)\bar{\alpha}_\textrm{atom}(i\omega)d\omega$,
where ${\alpha}_{\textrm{atom/ion}}(i\omega)$ is the dynamic polarizability of the atom/ion at imaginary frequency, and the
average polarizability and polarizability anisotropy are given by $\bar{\alpha}=(\alpha_\parallel+2\alpha_\perp)/3$ and $\Delta\alpha=\alpha_\parallel-\alpha_\perp$. Here, $\alpha_\parallel$ and $\alpha_\perp$ are polarizability tensor components parallel and perpendicular to the internuclear axis of the molecular ion.

The static electric dipole and quadrupole polarizabilities of the atoms ($\alpha_\textrm{Rb}=319.5\,$a.u., $\alpha_\textrm{Sr}=199.2\,$a.u., $\beta_\textrm{Rb}=6578\,$a.u., $\beta_\textrm{Sr}=4551\,$a.u.) and the permanent electric dipole and quadrupole moments of the molecular ion ($d_\textrm{ion}=-0.66\,$a.u., $\Theta_\textrm{ion}=-5.48\,$a.u.) are calculated using the CCSD(T) and finite field methods.  
The values of the dynamic polarizability at imaginary frequency for Rb 
and Sr are taken from Ref.~\cite{DerevienkoADNDT10}, whereas the value 
for CN$^-$ is obtained using the polarization propagator within the 
coupled cluster method~\cite{MoszynskiCCCC05,KoronaMP06}.

Fig.~\ref{fig:potential} shows the leading Legendre components of the interaction potentials for CN$^-$+Rb and CN$^-$+Sr. We estimate the uncertainty of the obtained interaction potentials to be around 2\%, which corresponds to $10^{-3}$~a.u.\ at the global minimum.  The leading long-range isotropic and anisotropic induction and dispersion interaction coefficients are listed in Table~\ref{tab:Cn}. The long-range interaction potentials computed according to Eq.~\eqref{eq:En} are used for distances larger than $r'=25\,$a.u. The agreement between the raw \textit{ab initio} data and the asymptotic expansion is on the order of 1\% at $r'=25\,$a.u. It is worth emphasizing that at this level of accuracy one can expect only qualitative results for ultracold collision energies. However, the goal of this Rapid Communication is to demonstrate that the order of magnitude for the predicted phenomena is within experimental reach. Furthermore, in the future the interaction potentials can be corrected using the scattering data from experiment, thereby allowing for a quantitative comparison.

\begin{figure}[t]
  \includegraphics[width=.9\columnwidth]{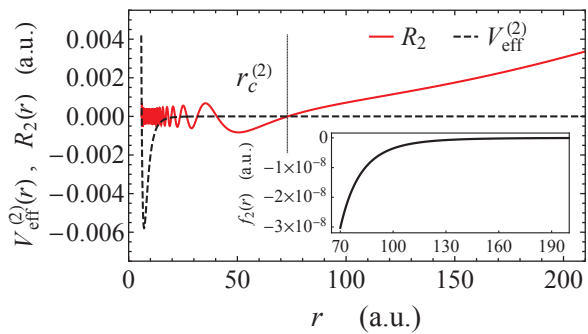}  
 \caption{ (Color online) The effective potential, $V_{\mathrm{eff}}^{(2)}(r)$, for CN$^-$ + $^{87}$Rb, and the corresponding zero-energy radial wave function, $R_2(r)$ (not normalized).  The cutoff radius is marked by a vertical line at $r^{(2)}_c = 72$ a.u.  The long-range behaviour of the corresponding Legendre component of the potential energy surface, $f_2(r) = -3615.7/r^6$, is shown in the inset. } 
 \label{fig-cutoff}
\end{figure}

{\it Quantum reflection regularized potentials.}
In the derivation of the Hamiltonian~\eqref{hamiltonian}, the impurity-boson interaction terms which are  bilinear in bosonic operators have been neglected. In the absence of resonant scattering \cite{Rath2013}, such terms are irrelevant to describe many-particle physics as they do not scale with the macroscopically large BEC density $n$ \cite{GirardeauPF61}. However, they are crucial for the  correct description of the underlying two-body scattering physics~\cite{Rath2013,Shchadilova2016}. As a consequence, when studying Eq.~\eqref{hamiltonian} the interaction potentials $f_\lambda(r)$ have to be properly regularized. Here we introduce a regularization scheme which is based on a pseudopotential approximation. 

In an ultracold  gas collisions take place at extremely low scattering energies. Therefore, bosons impinging on the molecular impurity cannot penetrate the short-range part of the   interaction potential and are quantum reflected at large distances \cite{friedrich2013scattering}. The corresponding scattering wave function is largely suppressed at small distances, implying that bosons are never probing the strong interactions at short range, which is not by default included  in the reduced Hamiltonian \eqref{hamiltonian}. 

Effectively, the suppression of boson density at small distances from the molecular ion can be modelled by introducing a short-distance cutoff in the interaction potentials. In order to determine this cutoff,  we solve the scattering problem of an atom interacting with the impurity. Let us first consider the case of a predominantly $s-$wave symmetric potential (i.e.~only $f_\lambda$ with $\lambda=0$ is nonzero). From the solution of the radial  Schr\"odinger equation we determine the outermost node of the scattering wave function and promote this position to the short distance cutoff $r_c^{(0)}$.  Such a construction effectively amounts to introducing a hard wall boundary condition at the node position $r_c^{(0)}$ which keeps the asymptotic scattering wave function intact. However, most importantly, by using this procedure one captures correctly the suppression of the wave function at distances $r<r_c^{(0)}$. 

The bosons, which form a BEC, are initially in their ground state. Consequently, when scattering off the impurity they impinge on the molecular ion predominately in their zero-angular momentum state. Since the scattering potential of a molecular ion is not spherically symmetric, it can induce angular momentum changing collisions so that bosons are scattered to  finite angular momentum $\lambda$ as induced by the potential contributions $f_\lambda(r)$. This fact can be used to define a semianalytic method to determine the short-distance cutoff $r_c^{(\lambda)}$ for each $\lambda$-channel of the potential, $f_\lambda(r)$, separately. For the many-body problem of BEC atoms the most relevant  scattering process is bosons which scattered from angular momentum zero to finite $\lambda$. Consequently, the outgoing atoms will be subject to the associated centrifugal barrier and we may determine the short distance cutoff $r_c^{(\lambda)}$  from the solution of the radial Schr\"odinger equation with the effective potential, $V^{(\lambda)}_{\mathrm{eff}}(r) = f_\lambda(r) + \frac{\lambda(\lambda+1)}{ 2 \mu r^2}$, evaluated at  zero collision energy as appropriate for ultracold experiments. Here $\mu$ is the reduced mass of the colliding pair.

 The procedure is illustrated in Fig.~\ref{fig-cutoff} for the case of CN$^-$ + $^{87}$Rb with $\lambda=2$. One can see that  the magnitude of the radial wavefunction, $R_\lambda(r)$, and therefore the corresponding probability amplitude, $|R_\lambda(r)|^2$,   is indeed vanishingly small at distances smaller than the position of the last node of the corresponding wavefunction. Introducing  short distance cutoffs in a similar way for all $\lambda$-channels, the Fourier components of the potential defined in Eq.~\eqref{potential_FT} can be replaced by the renormalized pseudo potentials
\begin{equation}
\tilde{U}_\lambda(k) =  \sqrt{\frac{32 \pi n k^2 {\varepsilon}_k}{{\omega}_k (2\lambda+1)^2}} {\int\limits_{ r_c^{(\lambda)}}}^\infty dr ~ r^2 f_\lambda(r) j_\lambda(kr), \label{cutoff}
\end{equation} 
which then serve as an input to the calculation of the angulon spectral function. For CN$^-$ + $^{87}$Rb and for zero collision energy, the cutoff radii are evaluated to $r^{(\lambda)}_c = 2150, 145, \text{and } 72$ a.u., for $\lambda =0,1,  \text{and } 2$ respectively. For CN$^-$ + $^{84}$Sr, on the other hand, those are given by $1000, 105,  \text{and }  45$ a.u.

It is worth mentioning that cutoff dependencies and the related renormalization of interactions poses a well known problem for  calculations of the Fr\"ohlich polaron properties in ultracold atomic systems~\cite{GrusdtCourse15}.   Since in the limit of isotropic interaction the Hamiltonian \eqref{hamiltonian} reduces to the Fr\"ohlich Hamiltonian, our regularization procedure can be used as a novel way to derive the effective coupling constants for Fr\"ohlich Hamiltonians from \textit{ab initio} potentials.

\begin{figure}[t]
 \includegraphics[width=.9\columnwidth]{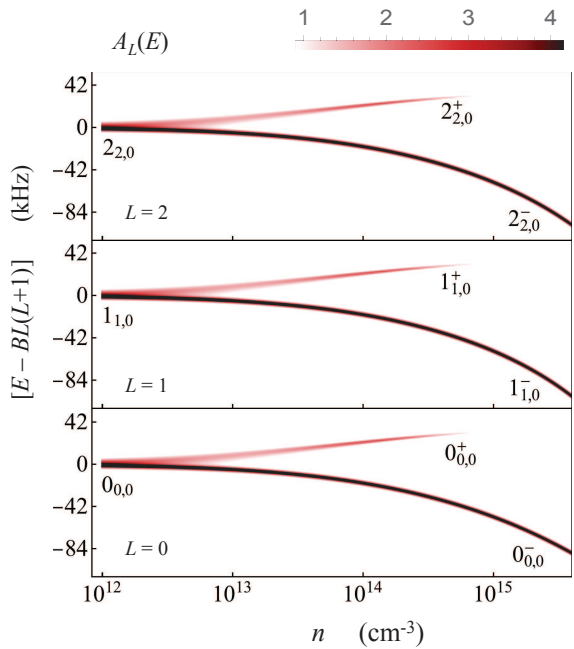}  
 \caption{(Color online) The angulon spectral function, $A_L(E)$, for CN$^-$ in a BEC of $^{87}$Rb, plotted as a function of the BEC density, $n$, on the logarithmic scale, and energy $E - B L(L+1)$. Only the three lowest  angulon states corresponding to total angular momenta, $L=0,1,2$, are shown. The lines were  artificially  broadened for better visibility.} \label{fig:spectral_function}
\end{figure}

{\it Rotational spectrum of CN$^-$ ion in $^{87}$Rb and $^{84}$Sr BECs.} In this section we employ the ansatz \eqref{ansatz} to evaluate the rotational spectrum of CN$^-$ in a BEC of $^{87}$Rb and $^{84}$Sr. The rotational constant of the CN$^-$ ion is given by $B =  2\pi \times 56$ GHz~\cite{Gottlieb07}; and  the scattering lengths for Rb--Rb and Sr--Sr  are given by $a_{bb} = 99$~a.u.~\cite{KempenPRL2002}, and $123$ a.u.\ \cite{KillianPRL2009}, respectively. Next, using the   effective interaction potentials and cutoff radii obtained in the previous sections, we calculate the spectral function $A_L(E)$, which is shown in Fig.~\ref{fig:spectral_function} for  CN$^-$ in a BEC of $^{87}$Rb of   typical densities between $10^{12}$ and $5\times10^{15}$~cm$^{-3}$. Qualitatively, the spectral function for CN$^-$ in a BEC of $^{84}$Sr is quite similar to Fig.~\ref{fig:spectral_function}, and is therefore not shown. In the density plot of Fig.~\ref{fig:spectral_function}, the dark shade (i.e.\ a peak in $A_L (E)$) corresponds to states with large quasiparticle weights and long lifetimes, while broader and lighter peaks correspond to unstable states. Following the notation of Ref.~\cite{SchmidtLem15}, the levels are labeled as $L_{j, \Lambda}$, where $L$ is the total angular momentum of the entire system, which is a good quantum number. The approximately conserved quantities, $j$ and $\Lambda$, on the other hand, give the value of angular momentum in the impurity and the bath, respectively. 
 The angulon spectrum reveals two  salient features arising due to the spherically-symmetric part of the molecule-BEC interaction. First is  the  ``polaron shift'': a uniform decrease in energy of all the angulon states with increasing density $n$.  Second is the ``many-body induced fine structure of first kind'', previously discussed in Refs.~\cite{SchmidtLem15,LemSchmidtChapter}. There, the spherically-symmetric part of the ion-atom interaction leads to the creation of spherically symmetric ($\lambda=0$) phonons, and the corresponding metastable excitation on top of the stable angulon state, see Fig.~\ref{fig:spectral_function}.  

\begin{figure}[tb]
  \includegraphics[width=.85\columnwidth]{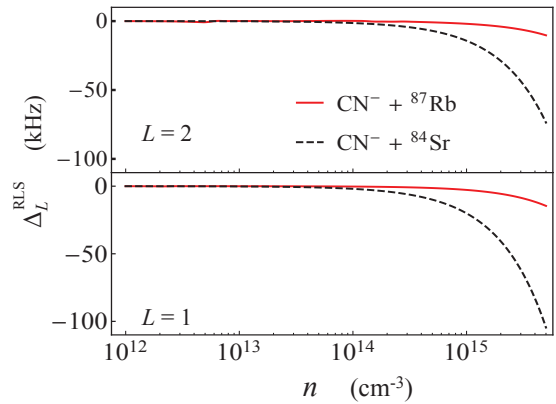}  
 \caption{(Color online) Differential rotational Lamb shifts for CN$^-$+$^{87}$Rb (red solid curve) and CN$^-$+$^{84}$Sr (black dashed curve), as a function of the BEC density, $n$, on a logarithmic scale.}
 \label{fig:RLS}
\end{figure}

The most interesting contribution, however, is the one arising due to the anisotropic molecule--BEC interactions leading to the striking angulon quasiparticle effects such as the rotational Lamb shift which is the rotational analog of a polaron effective mass and has a similar origin as the recently observed vibrational analog of the Lamb shift~\cite{oberthaler}.  In order to quantify such a contribution, we evaluate the differential rotational  Lamb shifts, as defined by $\Delta_L^{\mathrm{RLS}}(n)  = [E_L(n) - E_0(n)] - [E_L(0) - E_0(0)]$. The latter represents a difference in the self-energy acquired by different angular momentum states due to their interactions with a field of phonons in the BEC. The density dependence of the Lamb shifts is shown in Fig. \ref{fig:RLS} for CN$^-$  in both $^{87}$Rb and $^{84}$Sr. One can see that  the differential rotational Lamb-shifts for both   systems are on the order of a few kHz and therefore can be detected in modern experiments on cold ions. Furthermore, the value of $\Delta_L^{\mathrm{RLS}}$ is substantially larger for CN$^-$+$^{84}$Sr compared to that  for CN$^-$+$^{87}$Rb, due to the stronger anisotropy of the two-body interactions in the former case.

{\it Conclusions.}   We have shown that previously predicted physics of a rotating impurity interacting in a many-body bath~\cite{SchmidtLem15, SchmidtLem16, LemSchmidtChapter} can be studied in modern experiments on cold molecular ions immersed in a BEC. In particular, the rotational Lamb shifts arising due to the angular-momentum-dependent interactions between the molecular ion and the BEC were evaluated to be on the order of a few kHz; the predicted effects are not expected to  change substantially for the case of positive molecular ions. While the present theory does not include micromotion, which takes place in Paul traps~\cite{KrychPRA15}, the latter can be accounted for by incorporating an additional, translational degree of freedom for the impurity into our theory. In addition, including translational degrees of freedom will pave the way to studying rotationally inelastic collisions of molecular ions with a Bose-Einstein condensate~\cite{MidyaPreparation}.  Furthermore, the technique adopted here to determine the cut-off for potential energy surfaces can be used as a new renormalization procedure to obtain effective coupling constants for polaron models from \textit{ab initio} potentials. Finally we note that so far the formation of angulons has been studied only for bosonic baths with contact interactions  in three spatial dimensions. As for translational polarons \cite{prok2008,SchirotzekPRL09,schmidt_excitation_2011}, we expect the behavior of angulons to change dramatically when interacting with a fermionic bath \cite{anderson1967},  as well as with a bath featuring long-range electrostatic~\cite{LahayePfauRPP2009, LarzNJP15, LemKreDoyKais13} or laser-induced~\cite{LemeshkoPRA11Optical, LemFri11OpticalLong} interactions, or  whose motion is  constrained to reduced dimensions \cite{Schmidt2011a,KoschorreckNature12}.

{\it Acknowledgments.} The work was supported by the NSF through a grant for the Institute for Theoretical Atomic, Molecular, and Optical Physics at Harvard University and the Smithsonian Astrophysical Observatory. B.~M. acknowledges financial support received from the People Programme (Marie Curie Actions) of the European Union's Seventh Framework Programme (FP7/2007-2013) under REA grant agreement No. [291734].   M.~T. acknowledges support from the EU Marie Curie COFUND action (ICFOnest), the EU Grants ERC AdG OSYRIS, FP7 SIQS and EQuaM, FETPROACT QUIC, the Spanish Ministry Grants FOQUS (FIS2013-46768-P) and Severo Ochoa (SEV-2015-0522), Generalitat de Catalunya (SGR 874), Fundacio Cellex, the National Science Centre (2015/19/D/ST4/02173), and the PL-Grid Infrastructure.

\bibliography{ions.bib}

\begin{thebibliography}{88}%
\makeatletter
\providecommand \@ifxundefined [1]{%
 \@ifx{#1\undefined}
}%
\providecommand \@ifnum [1]{%
 \ifnum #1\expandafter \@firstoftwo
 \else \expandafter \@secondoftwo
 \fi
}%
\providecommand \@ifx [1]{%
 \ifx #1\expandafter \@firstoftwo
 \else \expandafter \@secondoftwo
 \fi
}%
\providecommand \natexlab [1]{#1}%
\providecommand \enquote  [1]{``#1''}%
\providecommand \bibnamefont  [1]{#1}%
\providecommand \bibfnamefont [1]{#1}%
\providecommand \citenamefont [1]{#1}%
\providecommand \href@noop [0]{\@secondoftwo}%
\providecommand \href [0]{\begingroup \@sanitize@url \@href}%
\providecommand \@href[1]{\@@startlink{#1}\@@href}%
\providecommand \@@href[1]{\endgroup#1\@@endlink}%
\providecommand \@sanitize@url [0]{\catcode `\\12\catcode `\$12\catcode
  `\&12\catcode `\#12\catcode `\^12\catcode `\_12\catcode `\%12\relax}%
\providecommand \@@startlink[1]{}%
\providecommand \@@endlink[0]{}%
\providecommand \url  [0]{\begingroup\@sanitize@url \@url }%
\providecommand \@url [1]{\endgroup\@href {#1}{\urlprefix }}%
\providecommand \urlprefix  [0]{URL }%
\providecommand \Eprint [0]{\href }%
\providecommand \doibase [0]{http://dx.doi.org/}%
\providecommand \selectlanguage [0]{\@gobble}%
\providecommand \bibinfo  [0]{\@secondoftwo}%
\providecommand \bibfield  [0]{\@secondoftwo}%
\providecommand \translation [1]{[#1]}%
\providecommand \BibitemOpen [0]{}%
\providecommand \bibitemStop [0]{}%
\providecommand \bibitemNoStop [0]{.\EOS\space}%
\providecommand \EOS [0]{\spacefactor3000\relax}%
\providecommand \BibitemShut  [1]{\csname bibitem#1\endcsname}%
\let\auto@bib@innerbib\@empty
\bibitem [{\citenamefont {Willitsch}(2012)}]{WillitschIRPC12}%
  \BibitemOpen
  \bibfield  {author} {\bibinfo {author} {\bibfnamefont {S.}~\bibnamefont
  {Willitsch}},\ }\href {\doibase 10.1080/0144235X.2012.667221} {\bibfield
  {journal} {\bibinfo  {journal} {Int. Rev. Phys. Chem.}\ }\textbf {\bibinfo
  {volume} {31}},\ \bibinfo {pages} {175} (\bibinfo {year} {2012})}\BibitemShut
  {NoStop}%
\bibitem [{\citenamefont {Willitsch}\ \emph {et~al.}(2008)\citenamefont
  {Willitsch}, \citenamefont {Bell}, \citenamefont {Gingell},\ and\
  \citenamefont {Softley}}]{WillitschPCCP2008}%
  \BibitemOpen
  \bibfield  {author} {\bibinfo {author} {\bibfnamefont {S.}~\bibnamefont
  {Willitsch}}, \bibinfo {author} {\bibfnamefont {M.~T.}\ \bibnamefont {Bell}},
  \bibinfo {author} {\bibfnamefont {A.~D.}\ \bibnamefont {Gingell}}, \ and\
  \bibinfo {author} {\bibfnamefont {T.~P.}\ \bibnamefont {Softley}},\ }\href
  {\doibase 10.1039/B813408C} {\bibfield  {journal} {\bibinfo  {journal} {Phys.
  Chem. Chem. Phys.}\ }\textbf {\bibinfo {volume} {10}},\ \bibinfo {pages}
  {7200} (\bibinfo {year} {2008})}\BibitemShut {NoStop}%
\bibitem [{\citenamefont {Lemeshko}\ \emph {et~al.}(2013)\citenamefont
  {Lemeshko}, \citenamefont {Krems}, \citenamefont {Doyle},\ and\ \citenamefont
  {Kais}}]{LemKreDoyKais13}%
  \BibitemOpen
  \bibfield  {author} {\bibinfo {author} {\bibfnamefont {M.}~\bibnamefont
  {Lemeshko}}, \bibinfo {author} {\bibfnamefont {R.}~\bibnamefont {Krems}},
  \bibinfo {author} {\bibfnamefont {J.}~\bibnamefont {Doyle}}, \ and\ \bibinfo
  {author} {\bibfnamefont {S.}~\bibnamefont {Kais}},\ }\href {\doibase
  10.1080/00268976.2013.813595} {\bibfield  {journal} {\bibinfo  {journal}
  {Mol. Phys.}\ }\textbf {\bibinfo {volume} {111}},\ \bibinfo {pages} {1648}
  (\bibinfo {year} {2013})}\BibitemShut {NoStop}%
\bibitem [{\citenamefont {Krems}\ \emph {et~al.}(2009)\citenamefont {Krems},
  \citenamefont {Stwalley},\ and\ \citenamefont
  {Friedrich}}]{KreStwFrieColdMol}%
  \BibitemOpen
  \bibinfo {editor} {\bibfnamefont {R.~V.}\ \bibnamefont {Krems}}, \bibinfo
  {editor} {\bibfnamefont {W.~C.}\ \bibnamefont {Stwalley}}, \ and\ \bibinfo
  {editor} {\bibfnamefont {B.}~\bibnamefont {Friedrich}},\ eds.,\ \href@noop {}
  {\emph {\bibinfo {title} {Cold molecules: theory, experiment,
  applications}}}\ (\bibinfo  {publisher} {Taylor\&Francis/CRC, Boca Raton,
  FL},\ \bibinfo {year} {2009})\BibitemShut {NoStop}%
\bibitem [{\citenamefont {Staanum}\ \emph {et~al.}(2010)\citenamefont
  {Staanum}, \citenamefont {H{\o}jbjerre}, \citenamefont {Skyt}, \citenamefont
  {Hansen},\ and\ \citenamefont {Drewsen}}]{StaanumNatPhys10}%
  \BibitemOpen
  \bibfield  {author} {\bibinfo {author} {\bibfnamefont {P.~F.}\ \bibnamefont
  {Staanum}}, \bibinfo {author} {\bibfnamefont {K.}~\bibnamefont
  {H{\o}jbjerre}}, \bibinfo {author} {\bibfnamefont {P.~S.}\ \bibnamefont
  {Skyt}}, \bibinfo {author} {\bibfnamefont {A.~K.}\ \bibnamefont {Hansen}}, \
  and\ \bibinfo {author} {\bibfnamefont {M.}~\bibnamefont {Drewsen}},\ }\href
  {\doibase 10.1038/nphys1604} {\bibfield  {journal} {\bibinfo  {journal} {Nat.
  Phys.}\ }\textbf {\bibinfo {volume} {6}},\ \bibinfo {pages} {271} (\bibinfo
  {year} {2010})}\BibitemShut {NoStop}%
\bibitem [{\citenamefont {Schneider}\ \emph {et~al.}(2010)\citenamefont
  {Schneider}, \citenamefont {Roth}, \citenamefont {Duncker}, \citenamefont
  {Ernsting},\ and\ \citenamefont {Schiller}}]{Schneider2010}%
  \BibitemOpen
  \bibfield  {author} {\bibinfo {author} {\bibfnamefont {T.}~\bibnamefont
  {Schneider}}, \bibinfo {author} {\bibfnamefont {B.}~\bibnamefont {Roth}},
  \bibinfo {author} {\bibfnamefont {H.}~\bibnamefont {Duncker}}, \bibinfo
  {author} {\bibfnamefont {I.}~\bibnamefont {Ernsting}}, \ and\ \bibinfo
  {author} {\bibfnamefont {S.}~\bibnamefont {Schiller}},\ }\href {\doibase
  10.1038/nphys1605} {\bibfield  {journal} {\bibinfo  {journal} {Nat. Phys.}\
  }\textbf {\bibinfo {volume} {6}},\ \bibinfo {pages} {275} (\bibinfo {year}
  {2010})}\BibitemShut {NoStop}%
\bibitem [{\citenamefont {Lien}\ \emph {et~al.}(2014)\citenamefont {Lien},
  \citenamefont {Seck}, \citenamefont {Lin}, \citenamefont {Nguyen},
  \citenamefont {Tabor},\ and\ \citenamefont {Odom}}]{Lien2014}%
  \BibitemOpen
  \bibfield  {author} {\bibinfo {author} {\bibfnamefont {C.}~\bibnamefont
  {Lien}}, \bibinfo {author} {\bibfnamefont {C.~M.}\ \bibnamefont {Seck}},
  \bibinfo {author} {\bibfnamefont {Y.}~\bibnamefont {Lin}}, \bibinfo {author}
  {\bibfnamefont {J.~H.~V.}\ \bibnamefont {Nguyen}}, \bibinfo {author}
  {\bibfnamefont {D.~A.}\ \bibnamefont {Tabor}}, \ and\ \bibinfo {author}
  {\bibfnamefont {B.~C.}\ \bibnamefont {Odom}},\ }\href {\doibase
  10.1038/ncomms5783} {\bibfield  {journal} {\bibinfo  {journal} {Nat.
  Commun.}\ }\textbf {\bibinfo {volume} {5}},\ \bibinfo {pages} {4783}
  (\bibinfo {year} {2014})}\BibitemShut {NoStop}%
\bibitem [{\citenamefont {Tong}\ \emph {et~al.}(2010)\citenamefont {Tong},
  \citenamefont {Winney},\ and\ \citenamefont {Willitsch}}]{TongPRL10}%
  \BibitemOpen
  \bibfield  {author} {\bibinfo {author} {\bibfnamefont {X.}~\bibnamefont
  {Tong}}, \bibinfo {author} {\bibfnamefont {A.~H.}\ \bibnamefont {Winney}}, \
  and\ \bibinfo {author} {\bibfnamefont {S.}~\bibnamefont {Willitsch}},\ }\href
  {\doibase 10.1103/PhysRevLett.105.143001} {\bibfield  {journal} {\bibinfo
  {journal} {Phys. Rev. Lett.}\ }\textbf {\bibinfo {volume} {105}},\ \bibinfo
  {pages} {143001} (\bibinfo {year} {2010})}\BibitemShut {NoStop}%
\bibitem [{\citenamefont {Mur-Petiti}\ \emph {et~al.}(2013)\citenamefont
  {Mur-Petiti}, \citenamefont {P\'erez-R\'ios}, \citenamefont
  {Campos-Mart\'inez}, \citenamefont {Hern\'andez}, \citenamefont {Willitsch},\
  and\ \citenamefont {Garc\'ia-Ripoll}}]{Petit2013}%
  \BibitemOpen
  \bibfield  {author} {\bibinfo {author} {\bibfnamefont {J.}~\bibnamefont
  {Mur-Petiti}}, \bibinfo {author} {\bibfnamefont {J.}~\bibnamefont
  {P\'erez-R\'ios}}, \bibinfo {author} {\bibfnamefont {J.}~\bibnamefont
  {Campos-Mart\'inez}}, \bibinfo {author} {\bibfnamefont {M.~I.}\ \bibnamefont
  {Hern\'andez}}, \bibinfo {author} {\bibfnamefont {S.}~\bibnamefont
  {Willitsch}}, \ and\ \bibinfo {author} {\bibfnamefont {J.~J.}\ \bibnamefont
  {Garc\'ia-Ripoll}},\ }\href@noop {} {\emph {\bibinfo {title} {in Architecture
  and Design of Molecule Logic Gates and Atom Circuits}}},\ edited by\ \bibinfo
  {editor} {\bibfnamefont {N.}~\bibnamefont {Lorente}}\ and\ \bibinfo {editor}
  {\bibfnamefont {C.}~\bibnamefont {Joachim}}\ (\bibinfo  {publisher} {Springer
  Berlin Heidelberg},\ \bibinfo {year} {2013})\ pp.\ \bibinfo {pages}
  {267--277}\BibitemShut {NoStop}%
\bibitem [{\citenamefont {Shi}\ \emph {et~al.}(2013)\citenamefont {Shi},
  \citenamefont {Herskind}, \citenamefont {Drewsen},\ and\ \citenamefont
  {Chuang}}]{ShiNJP13}%
  \BibitemOpen
  \bibfield  {author} {\bibinfo {author} {\bibfnamefont {M.}~\bibnamefont
  {Shi}}, \bibinfo {author} {\bibfnamefont {P.~F.}\ \bibnamefont {Herskind}},
  \bibinfo {author} {\bibfnamefont {M.}~\bibnamefont {Drewsen}}, \ and\
  \bibinfo {author} {\bibfnamefont {I.~L.}\ \bibnamefont {Chuang}},\ }\href
  {\doibase 10.1088/1367-2630/15/11/113019} {\bibfield  {journal} {\bibinfo
  {journal} {New J. Phys.}\ }\textbf {\bibinfo {volume} {15}},\ \bibinfo
  {pages} {113019} (\bibinfo {year} {2013})}\BibitemShut {NoStop}%
\bibitem [{\citenamefont {Larsson}\ \emph {et~al.}(2012)\citenamefont
  {Larsson}, \citenamefont {Geppert},\ and\ \citenamefont
  {Nyman}}]{LarssonRPP12}%
  \BibitemOpen
  \bibfield  {author} {\bibinfo {author} {\bibfnamefont {M.}~\bibnamefont
  {Larsson}}, \bibinfo {author} {\bibfnamefont {W.~D.}\ \bibnamefont
  {Geppert}}, \ and\ \bibinfo {author} {\bibfnamefont {G.}~\bibnamefont
  {Nyman}},\ }\href {\doibase 10.1088/0034-4885/75/6/066901} {\bibfield
  {journal} {\bibinfo  {journal} {Rep. Prog. Phys.}\ }\textbf {\bibinfo
  {volume} {75}},\ \bibinfo {pages} {066901} (\bibinfo {year}
  {2012})}\BibitemShut {NoStop}%
\bibitem [{\citenamefont {Fortenberry}(2015)}]{FortenberryJPCA15}%
  \BibitemOpen
  \bibfield  {author} {\bibinfo {author} {\bibfnamefont {R.~C.}\ \bibnamefont
  {Fortenberry}},\ }\href {\doibase 10.1021/acs.jpca.5b05056} {\bibfield
  {journal} {\bibinfo  {journal} {J. Phys. Chem. A}\ }\textbf {\bibinfo
  {volume} {119}},\ \bibinfo {pages} {9941} (\bibinfo {year}
  {2015})}\BibitemShut {NoStop}%
\bibitem [{\citenamefont {Hall}\ and\ \citenamefont
  {Willitsch}(2012)}]{HallPRL12}%
  \BibitemOpen
  \bibfield  {author} {\bibinfo {author} {\bibfnamefont {F.~H.~J.}\
  \bibnamefont {Hall}}\ and\ \bibinfo {author} {\bibfnamefont {S.}~\bibnamefont
  {Willitsch}},\ }\href {\doibase 10.1103/PhysRevLett.109.233202} {\bibfield
  {journal} {\bibinfo  {journal} {Phys. Rev. Lett.}\ }\textbf {\bibinfo
  {volume} {109}},\ \bibinfo {pages} {233202} (\bibinfo {year}
  {2012})}\BibitemShut {NoStop}%
\bibitem [{\citenamefont {H{\o}jbjerre}\ \emph {et~al.}(2008)\citenamefont
  {H{\o}jbjerre}, \citenamefont {Offenberg}, \citenamefont {Bisgaard},
  \citenamefont {Stapelfeldt}, \citenamefont {Staanum}, \citenamefont
  {Mortensen},\ and\ \citenamefont {Drewsen}}]{HojbjerrePRA08}%
  \BibitemOpen
  \bibfield  {author} {\bibinfo {author} {\bibfnamefont {K.}~\bibnamefont
  {H{\o}jbjerre}}, \bibinfo {author} {\bibfnamefont {D.}~\bibnamefont
  {Offenberg}}, \bibinfo {author} {\bibfnamefont {C.}~\bibnamefont {Bisgaard}},
  \bibinfo {author} {\bibfnamefont {H.}~\bibnamefont {Stapelfeldt}}, \bibinfo
  {author} {\bibfnamefont {P.}~\bibnamefont {Staanum}}, \bibinfo {author}
  {\bibfnamefont {A.}~\bibnamefont {Mortensen}}, \ and\ \bibinfo {author}
  {\bibfnamefont {M.}~\bibnamefont {Drewsen}},\ }\href {\doibase
  10.1103/PhysRevA.77.030702} {\bibfield  {journal} {\bibinfo  {journal} {Phys.
  Rev. A}\ }\textbf {\bibinfo {volume} {77}},\ \bibinfo {pages} {030702(R)}
  (\bibinfo {year} {2008})}\BibitemShut {NoStop}%
\bibitem [{\citenamefont {Bressel}\ \emph {et~al.}(2012)\citenamefont
  {Bressel}, \citenamefont {Borodin}, \citenamefont {Shen}, \citenamefont
  {Hansen}, \citenamefont {Ernsting},\ and\ \citenamefont
  {Schiller}}]{BresselPRL12}%
  \BibitemOpen
  \bibfield  {author} {\bibinfo {author} {\bibfnamefont {U.}~\bibnamefont
  {Bressel}}, \bibinfo {author} {\bibfnamefont {A.}~\bibnamefont {Borodin}},
  \bibinfo {author} {\bibfnamefont {J.}~\bibnamefont {Shen}}, \bibinfo {author}
  {\bibfnamefont {M.}~\bibnamefont {Hansen}}, \bibinfo {author} {\bibfnamefont
  {I.}~\bibnamefont {Ernsting}}, \ and\ \bibinfo {author} {\bibfnamefont
  {S.}~\bibnamefont {Schiller}},\ }\href {\doibase
  10.1103/PhysRevLett.108.183003} {\bibfield  {journal} {\bibinfo  {journal}
  {Phys. Rev. Lett.}\ }\textbf {\bibinfo {volume} {108}},\ \bibinfo {pages}
  {183003} (\bibinfo {year} {2012})}\BibitemShut {NoStop}%
\bibitem [{\citenamefont {Germann}\ \emph {et~al.}(2014)\citenamefont
  {Germann}, \citenamefont {Tong},\ and\ \citenamefont
  {Willitsch}}]{GerToWillNatPhys2014}%
  \BibitemOpen
  \bibfield  {author} {\bibinfo {author} {\bibfnamefont {M.}~\bibnamefont
  {Germann}}, \bibinfo {author} {\bibfnamefont {X.}~\bibnamefont {Tong}}, \
  and\ \bibinfo {author} {\bibfnamefont {S.}~\bibnamefont {Willitsch}},\ }\href
  {\doibase 10.1038/nphys3085} {\bibfield  {journal} {\bibinfo  {journal} {Nat.
  Phys.}\ }\textbf {\bibinfo {volume} {10}},\ \bibinfo {pages} {820} (\bibinfo
  {year} {2014})}\BibitemShut {NoStop}%
\bibitem [{\citenamefont {Borschevsky}\ \emph {et~al.}(2012)\citenamefont
  {Borschevsky}, \citenamefont {Ilia\v{s}}, \citenamefont {Dzuba},
  \citenamefont {Beloy}, \citenamefont {Flambaum},\ and\ \citenamefont
  {Schwerdtfeger}}]{BorschevskyPRA2012}%
  \BibitemOpen
  \bibfield  {author} {\bibinfo {author} {\bibfnamefont {A.}~\bibnamefont
  {Borschevsky}}, \bibinfo {author} {\bibfnamefont {M.}~\bibnamefont
  {Ilia\v{s}}}, \bibinfo {author} {\bibfnamefont {V.~A.}\ \bibnamefont
  {Dzuba}}, \bibinfo {author} {\bibfnamefont {K.}~\bibnamefont {Beloy}},
  \bibinfo {author} {\bibfnamefont {V.~V.}\ \bibnamefont {Flambaum}}, \ and\
  \bibinfo {author} {\bibfnamefont {P.}~\bibnamefont {Schwerdtfeger}},\ }\href
  {\doibase 10.1103/PhysRevA.86.050501} {\bibfield  {journal} {\bibinfo
  {journal} {Phys. Rev. A}\ }\textbf {\bibinfo {volume} {86}},\ \bibinfo
  {pages} {050501(R)} (\bibinfo {year} {2012})}\BibitemShut {NoStop}%
\bibitem [{\citenamefont {Leanhardt}\ \emph {et~al.}(2011)\citenamefont
  {Leanhardt}, \citenamefont {Bohn}, \citenamefont {Loh}, \citenamefont
  {Maletinsky}, \citenamefont {Meyer}, \citenamefont {Sinclair}, \citenamefont
  {Stutz},\ and\ \citenamefont {Cornell}}]{LeanhardtJmolSpec11}%
  \BibitemOpen
  \bibfield  {author} {\bibinfo {author} {\bibfnamefont {A.~E.}\ \bibnamefont
  {Leanhardt}}, \bibinfo {author} {\bibfnamefont {J.~L.}\ \bibnamefont {Bohn}},
  \bibinfo {author} {\bibfnamefont {H.}~\bibnamefont {Loh}}, \bibinfo {author}
  {\bibfnamefont {P.}~\bibnamefont {Maletinsky}}, \bibinfo {author}
  {\bibfnamefont {E.~R.}\ \bibnamefont {Meyer}}, \bibinfo {author}
  {\bibfnamefont {L.~C.}\ \bibnamefont {Sinclair}}, \bibinfo {author}
  {\bibfnamefont {R.~P.}\ \bibnamefont {Stutz}}, \ and\ \bibinfo {author}
  {\bibfnamefont {E.~A.}\ \bibnamefont {Cornell}},\ }\href {\doibase
  10.1016/j.jms.2011.06.007} {\bibfield  {journal} {\bibinfo  {journal} {J.
  Mol. Spec.}\ }\textbf {\bibinfo {volume} {270}},\ \bibinfo {pages} {1}
  (\bibinfo {year} {2011})}\BibitemShut {NoStop}%
\bibitem [{\citenamefont {Loh}\ \emph {et~al.}(2013)\citenamefont {Loh},
  \citenamefont {Cossel}, \citenamefont {Grau}, \citenamefont {Ni},
  \citenamefont {Meyer}, \citenamefont {Bohn}, \citenamefont {Ye},\ and\
  \citenamefont {Cornell}}]{LohScience13}%
  \BibitemOpen
  \bibfield  {author} {\bibinfo {author} {\bibfnamefont {H.}~\bibnamefont
  {Loh}}, \bibinfo {author} {\bibfnamefont {K.~C.}\ \bibnamefont {Cossel}},
  \bibinfo {author} {\bibfnamefont {M.~C.}\ \bibnamefont {Grau}}, \bibinfo
  {author} {\bibfnamefont {K.-K.}\ \bibnamefont {Ni}}, \bibinfo {author}
  {\bibfnamefont {E.~R.}\ \bibnamefont {Meyer}}, \bibinfo {author}
  {\bibfnamefont {J.~L.}\ \bibnamefont {Bohn}}, \bibinfo {author}
  {\bibfnamefont {J.}~\bibnamefont {Ye}}, \ and\ \bibinfo {author}
  {\bibfnamefont {E.~A.}\ \bibnamefont {Cornell}},\ }\href {\doibase
  10.1126/science.1243683} {\bibfield  {journal} {\bibinfo  {journal}
  {Science}\ }\textbf {\bibinfo {volume} {342}},\ \bibinfo {pages} {1220}
  (\bibinfo {year} {2013})}\BibitemShut {NoStop}%
\bibitem [{\citenamefont {Wolf}\ \emph {et~al.}(2016)\citenamefont {Wolf},
  \citenamefont {Wan}, \citenamefont {Heip}, \citenamefont {Gebert},
  \citenamefont {Shi},\ and\ \citenamefont {Schmidt}}]{WolfNature2016}%
  \BibitemOpen
  \bibfield  {author} {\bibinfo {author} {\bibfnamefont {F.}~\bibnamefont
  {Wolf}}, \bibinfo {author} {\bibfnamefont {Y.}~\bibnamefont {Wan}}, \bibinfo
  {author} {\bibfnamefont {J.~C.}\ \bibnamefont {Heip}}, \bibinfo {author}
  {\bibfnamefont {F.}~\bibnamefont {Gebert}}, \bibinfo {author} {\bibfnamefont
  {C.}~\bibnamefont {Shi}}, \ and\ \bibinfo {author} {\bibfnamefont {P.~O.}\
  \bibnamefont {Schmidt}},\ }\href {\doibase 10.1038/nature16513} {\bibfield
  {journal} {\bibinfo  {journal} {Nature}\ }\textbf {\bibinfo {volume} {530}},\
  \bibinfo {pages} {457} (\bibinfo {year} {2016})}\BibitemShut {NoStop}%
\bibitem [{\citenamefont {Biesheuvel}\ \emph {et~al.}(2016)\citenamefont
  {Biesheuvel}, \citenamefont {Karr}, \citenamefont {Hilico}, \citenamefont
  {Eikema}, \citenamefont {Ubachs},\ and\ \citenamefont
  {Koelemeij}}]{BiesheuvelNatComm16}%
  \BibitemOpen
  \bibfield  {author} {\bibinfo {author} {\bibfnamefont {J.}~\bibnamefont
  {Biesheuvel}}, \bibinfo {author} {\bibfnamefont {J.~P.}\ \bibnamefont
  {Karr}}, \bibinfo {author} {\bibfnamefont {L.}~\bibnamefont {Hilico}},
  \bibinfo {author} {\bibfnamefont {K.~S.~E.}\ \bibnamefont {Eikema}}, \bibinfo
  {author} {\bibfnamefont {W.}~\bibnamefont {Ubachs}}, \ and\ \bibinfo {author}
  {\bibfnamefont {J.~C.~J.}\ \bibnamefont {Koelemeij}},\ }\href {\doibase
  10.1038/ncomms10385} {\bibfield  {journal} {\bibinfo  {journal} {Nat Commun}\
  }\textbf {\bibinfo {volume} {7}},\ \bibinfo {pages} {10385} (\bibinfo {year}
  {2016})}\BibitemShut {NoStop}%
\bibitem [{\citenamefont {Lewenstein}\ \emph {et~al.}(2012)\citenamefont
  {Lewenstein}, \citenamefont {Sanpera},\ and\ \citenamefont
  {Ahufinger}}]{LewensteinBook12}%
  \BibitemOpen
  \bibfield  {author} {\bibinfo {author} {\bibfnamefont {M.}~\bibnamefont
  {Lewenstein}}, \bibinfo {author} {\bibfnamefont {A.}~\bibnamefont {Sanpera}},
  \ and\ \bibinfo {author} {\bibfnamefont {V.}~\bibnamefont {Ahufinger}},\
  }\href@noop {} {\emph {\bibinfo {title} {Ultracold Atoms in Optical Lattices:
  Simulating quantum many-body systems}}}\ (\bibinfo  {publisher} {Oxford},\
  \bibinfo {year} {2012})\BibitemShut {NoStop}%
\bibitem [{\citenamefont {Carr}\ \emph {et~al.}(2009)\citenamefont {Carr},
  \citenamefont {DeMille}, \citenamefont {Krems},\ and\ \citenamefont
  {Ye}}]{Carr09}%
  \BibitemOpen
  \bibfield  {author} {\bibinfo {author} {\bibfnamefont {L.~D.}\ \bibnamefont
  {Carr}}, \bibinfo {author} {\bibfnamefont {D.}~\bibnamefont {DeMille}},
  \bibinfo {author} {\bibfnamefont {R.~V.}\ \bibnamefont {Krems}}, \ and\
  \bibinfo {author} {\bibfnamefont {J.}~\bibnamefont {Ye}},\ }\href {\doibase
  10.1088/1367-2630/11/5/055049} {\bibfield  {journal} {\bibinfo  {journal}
  {New J. Phys.}\ }\textbf {\bibinfo {volume} {11}},\ \bibinfo {pages} {055049}
  (\bibinfo {year} {2009})}\BibitemShut {NoStop}%
\bibitem [{\citenamefont {Jin}\ and\ \citenamefont {Ye}(2012)}]{JinYeCRev12}%
  \BibitemOpen
  \bibfield  {author} {\bibinfo {author} {\bibfnamefont {D.~S.}\ \bibnamefont
  {Jin}}\ and\ \bibinfo {author} {\bibfnamefont {J.}~\bibnamefont {Ye}},\
  }\href {\doibase 10.1021/cr300342x} {\bibfield  {journal} {\bibinfo
  {journal} {Chem. Rev.}\ }\textbf {\bibinfo {volume} {112}},\ \bibinfo {pages}
  {4801} (\bibinfo {year} {2012})}\BibitemShut {NoStop}%
\bibitem [{\citenamefont {Yan}\ \emph {et~al.}(2013)\citenamefont {Yan},
  \citenamefont {Moses}, \citenamefont {Gadway}, \citenamefont {Covey},
  \citenamefont {Hazzard}, \citenamefont {Rey}, \citenamefont {Jin},\ and\
  \citenamefont {Ye}}]{YanNature13}%
  \BibitemOpen
  \bibfield  {author} {\bibinfo {author} {\bibfnamefont {B.}~\bibnamefont
  {Yan}}, \bibinfo {author} {\bibfnamefont {S.~A.}\ \bibnamefont {Moses}},
  \bibinfo {author} {\bibfnamefont {B.}~\bibnamefont {Gadway}}, \bibinfo
  {author} {\bibfnamefont {J.~P.}\ \bibnamefont {Covey}}, \bibinfo {author}
  {\bibfnamefont {K.~R.~A.}\ \bibnamefont {Hazzard}}, \bibinfo {author}
  {\bibfnamefont {A.~M.}\ \bibnamefont {Rey}}, \bibinfo {author} {\bibfnamefont
  {D.~S.}\ \bibnamefont {Jin}}, \ and\ \bibinfo {author} {\bibfnamefont
  {J.}~\bibnamefont {Ye}},\ }\href {\doibase 10.1038/nature12483} {\bibfield
  {journal} {\bibinfo  {journal} {Nature}\ }\textbf {\bibinfo {volume} {501}},\
  \bibinfo {pages} {521} (\bibinfo {year} {2013})}\BibitemShut {NoStop}%
\bibitem [{\citenamefont {Stuhl}\ \emph {et~al.}(2014)\citenamefont {Stuhl},
  \citenamefont {Hummon},\ and\ \citenamefont {Ye}}]{YeARPC2014}%
  \BibitemOpen
  \bibfield  {author} {\bibinfo {author} {\bibfnamefont {B.~K.}\ \bibnamefont
  {Stuhl}}, \bibinfo {author} {\bibfnamefont {M.~T.}\ \bibnamefont {Hummon}}, \
  and\ \bibinfo {author} {\bibfnamefont {J.}~\bibnamefont {Ye}},\ }\href
  {\doibase 10.1146/annurev-physchem-040513-103744} {\bibfield  {journal}
  {\bibinfo  {journal} {Ann. Rev. Phys. Chem.}\ }\textbf {\bibinfo {volume}
  {65}},\ \bibinfo {pages} {501} (\bibinfo {year} {2014})}\BibitemShut
  {NoStop}%
\bibitem [{\citenamefont {Rellergert}\ \emph {et~al.}(2013)\citenamefont
  {Rellergert}, \citenamefont {Sullivan}, \citenamefont {Schowalter},
  \citenamefont {Kotochigova}, \citenamefont {Chen},\ and\ \citenamefont
  {Hudson}}]{RellergertNat13}%
  \BibitemOpen
  \bibfield  {author} {\bibinfo {author} {\bibfnamefont {W.~G.}\ \bibnamefont
  {Rellergert}}, \bibinfo {author} {\bibfnamefont {S.~T.}\ \bibnamefont
  {Sullivan}}, \bibinfo {author} {\bibfnamefont {S.~J.}\ \bibnamefont
  {Schowalter}}, \bibinfo {author} {\bibfnamefont {S.}~\bibnamefont
  {Kotochigova}}, \bibinfo {author} {\bibfnamefont {K.}~\bibnamefont {Chen}}, \
  and\ \bibinfo {author} {\bibfnamefont {E.~R.}\ \bibnamefont {Hudson}},\
  }\href {\doibase 10.1038/nature11937} {\bibfield  {journal} {\bibinfo
  {journal} {Nature}\ }\textbf {\bibinfo {volume} {495}},\ \bibinfo {pages}
  {490} (\bibinfo {year} {2013})}\BibitemShut {NoStop}%
\bibitem [{\citenamefont {Stoecklin}\ \emph {et~al.}(2016)\citenamefont
  {Stoecklin}, \citenamefont {Halvick}, \citenamefont {Gannouni}, \citenamefont
  {Hochlaf}, \citenamefont {Kotochigova},\ and\ \citenamefont
  {Hudson}}]{StoecklinNatComm16}%
  \BibitemOpen
  \bibfield  {author} {\bibinfo {author} {\bibfnamefont {T.}~\bibnamefont
  {Stoecklin}}, \bibinfo {author} {\bibfnamefont {P.}~\bibnamefont {Halvick}},
  \bibinfo {author} {\bibfnamefont {M.~A.}\ \bibnamefont {Gannouni}}, \bibinfo
  {author} {\bibfnamefont {M.}~\bibnamefont {Hochlaf}}, \bibinfo {author}
  {\bibfnamefont {S.}~\bibnamefont {Kotochigova}}, \ and\ \bibinfo {author}
  {\bibfnamefont {E.~R.}\ \bibnamefont {Hudson}},\ }\href {\doibase
  10.1038/ncomms11234} {\bibfield  {journal} {\bibinfo  {journal} {Nat.
  Commun.}\ }\textbf {\bibinfo {volume} {7}},\ \bibinfo {pages} {11234}
  (\bibinfo {year} {2016})}\BibitemShut {NoStop}%
\bibitem [{\citenamefont {Hauser}\ \emph {et~al.}(2015)\citenamefont {Hauser},
  \citenamefont {Lee}, \citenamefont {Carelli}, \citenamefont {Spieler},
  \citenamefont {Lakhmanskaya}, \citenamefont {Endres}, \citenamefont {Kumar},
  \citenamefont {Gianturco},\ and\ \citenamefont {Wester}}]{HauserNatPhys15}%
  \BibitemOpen
  \bibfield  {author} {\bibinfo {author} {\bibfnamefont {D.}~\bibnamefont
  {Hauser}}, \bibinfo {author} {\bibfnamefont {S.}~\bibnamefont {Lee}},
  \bibinfo {author} {\bibfnamefont {F.}~\bibnamefont {Carelli}}, \bibinfo
  {author} {\bibfnamefont {S.}~\bibnamefont {Spieler}}, \bibinfo {author}
  {\bibfnamefont {O.}~\bibnamefont {Lakhmanskaya}}, \bibinfo {author}
  {\bibfnamefont {E.~S.}\ \bibnamefont {Endres}}, \bibinfo {author}
  {\bibfnamefont {S.~S.}\ \bibnamefont {Kumar}}, \bibinfo {author}
  {\bibfnamefont {F.}~\bibnamefont {Gianturco}}, \ and\ \bibinfo {author}
  {\bibfnamefont {R.}~\bibnamefont {Wester}},\ }\href {\doibase
  10.1038/nphys3326} {\bibfield  {journal} {\bibinfo  {journal} {Nat. Phys.}\
  }\textbf {\bibinfo {volume} {11}},\ \bibinfo {pages} {467} (\bibinfo {year}
  {2015})}\BibitemShut {NoStop}%
\bibitem [{\citenamefont {Otto}\ \emph {et~al.}(2008)\citenamefont {Otto},
  \citenamefont {Mikosch}, \citenamefont {Trippel}, \citenamefont
  {Weidem\"uller},\ and\ \citenamefont {Wester}}]{OttoPRL08}%
  \BibitemOpen
  \bibfield  {author} {\bibinfo {author} {\bibfnamefont {R.}~\bibnamefont
  {Otto}}, \bibinfo {author} {\bibfnamefont {J.}~\bibnamefont {Mikosch}},
  \bibinfo {author} {\bibfnamefont {S.}~\bibnamefont {Trippel}}, \bibinfo
  {author} {\bibfnamefont {M.}~\bibnamefont {Weidem\"uller}}, \ and\ \bibinfo
  {author} {\bibfnamefont {R.}~\bibnamefont {Wester}},\ }\href {\doibase
  10.1103/PhysRevLett.101.063201} {\bibfield  {journal} {\bibinfo  {journal}
  {Phys. Rev. Lett.}\ }\textbf {\bibinfo {volume} {101}},\ \bibinfo {pages}
  {063201} (\bibinfo {year} {2008})}\BibitemShut {NoStop}%
\bibitem [{\citenamefont {Deiglmayr}\ \emph {et~al.}(2012)\citenamefont
  {Deiglmayr}, \citenamefont {G\"oritz}, \citenamefont {Best}, \citenamefont
  {Weidem\"uller},\ and\ \citenamefont {Wester}}]{DeiglmayrPRA12}%
  \BibitemOpen
  \bibfield  {author} {\bibinfo {author} {\bibfnamefont {J.}~\bibnamefont
  {Deiglmayr}}, \bibinfo {author} {\bibfnamefont {A.}~\bibnamefont {G\"oritz}},
  \bibinfo {author} {\bibfnamefont {T.}~\bibnamefont {Best}}, \bibinfo {author}
  {\bibfnamefont {M.}~\bibnamefont {Weidem\"uller}}, \ and\ \bibinfo {author}
  {\bibfnamefont {R.}~\bibnamefont {Wester}},\ }\href {\doibase
  10.1103/PhysRevA.86.043438} {\bibfield  {journal} {\bibinfo  {journal} {Phys.
  Rev. A}\ }\textbf {\bibinfo {volume} {86}},\ \bibinfo {pages} {043438}
  (\bibinfo {year} {2012})}\BibitemShut {NoStop}%
\bibitem [{\citenamefont {Schmidt}\ and\ \citenamefont
  {Lemeshko}(2015)}]{SchmidtLem15}%
  \BibitemOpen
  \bibfield  {author} {\bibinfo {author} {\bibfnamefont {R.}~\bibnamefont
  {Schmidt}}\ and\ \bibinfo {author} {\bibfnamefont {M.}~\bibnamefont
  {Lemeshko}},\ }\href {\doibase 10.1103/PhysRevLett.114.203001} {\bibfield
  {journal} {\bibinfo  {journal} {Phys. Rev. Lett.}\ }\textbf {\bibinfo
  {volume} {114}},\ \bibinfo {pages} {203001} (\bibinfo {year}
  {2015})}\BibitemShut {NoStop}%
\bibitem [{\citenamefont {Schmidt}\ and\ \citenamefont
  {Lemeshko}(2016)}]{SchmidtLem16}%
  \BibitemOpen
  \bibfield  {author} {\bibinfo {author} {\bibfnamefont {R.}~\bibnamefont
  {Schmidt}}\ and\ \bibinfo {author} {\bibfnamefont {M.}~\bibnamefont
  {Lemeshko}},\ }\href {\doibase 10.1103/PhysRevX.6.011012} {\bibfield
  {journal} {\bibinfo  {journal} {Phys. Rev. X}\ }\textbf {\bibinfo {volume}
  {6}},\ \bibinfo {pages} {011012} (\bibinfo {year} {2016})}\BibitemShut
  {NoStop}%
\bibitem [{\citenamefont {Lemeshko}\ and\ \citenamefont
  {Schmidt}(2016)}]{LemSchmidtChapter}%
  \BibitemOpen
  \bibfield  {author} {\bibinfo {author} {\bibfnamefont {M.}~\bibnamefont
  {Lemeshko}}\ and\ \bibinfo {author} {\bibfnamefont {R.}~\bibnamefont
  {Schmidt}},\ }\enquote {\bibinfo {title} {Molecular impurities interacting
  with a many-particle environment: from ultracold gases to helium
  nanodroplets},}\ in\ \href@noop {} {\emph {\bibinfo {booktitle} {Low Energy
  and Low Temperature Molecular Scattering}}},\ \bibinfo {editor} {edited by\
  \bibinfo {editor} {\bibfnamefont {A.}~\bibnamefont {Osterwalder}}\ and\
  \bibinfo {editor} {\bibfnamefont {O.}~\bibnamefont {Dulieu}}}\ (\bibinfo
  {publisher} {RSC},\ \bibinfo {year} {2016})\BibitemShut {NoStop}%
\bibitem [{\citenamefont {Landau}(1933)}]{LandauPolaron}%
  \BibitemOpen
  \bibfield  {author} {\bibinfo {author} {\bibfnamefont {L.~D.}\ \bibnamefont
  {Landau}},\ }\href@noop {} {\bibfield  {journal} {\bibinfo  {journal} {Phys.
  Z. Sowjetunion}\ }\textbf {\bibinfo {volume} {3}},\ \bibinfo {pages} {664}
  (\bibinfo {year} {1933})}\BibitemShut {NoStop}%
\bibitem [{\citenamefont {Fr{\"o}hlich}(1954)}]{FrohlichAdvPhys54}%
  \BibitemOpen
  \bibfield  {author} {\bibinfo {author} {\bibfnamefont {H.}~\bibnamefont
  {Fr{\"o}hlich}},\ }\href {\doibase 10.1080/00018735400101213} {\bibfield
  {journal} {\bibinfo  {journal} {Adv. Phys.}\ }\textbf {\bibinfo {volume}
  {3}},\ \bibinfo {pages} {325} (\bibinfo {year} {1954})}\BibitemShut {NoStop}%
\bibitem [{\citenamefont {Feynman}(1955)}]{FeynmanPR55}%
  \BibitemOpen
  \bibfield  {author} {\bibinfo {author} {\bibfnamefont {R.~P.}\ \bibnamefont
  {Feynman}},\ }\href {\doibase 10.1103/PhysRev.97.660} {\bibfield  {journal}
  {\bibinfo  {journal} {Phys. Rev.}\ }\textbf {\bibinfo {volume} {97}},\
  \bibinfo {pages} {660} (\bibinfo {year} {1955})}\BibitemShut {NoStop}%
\bibitem [{\citenamefont {Devreese}(2015)}]{Devreese15}%
  \BibitemOpen
  \bibfield  {author} {\bibinfo {author} {\bibfnamefont {J.~T.}\ \bibnamefont
  {Devreese}},\ }\href {http://arxiv.org/abs/1012.4576} {\bibfield  {journal}
  {\bibinfo  {journal} {arXiv:1012.4576v6}\ } (\bibinfo {year}
  {2015})}\BibitemShut {NoStop}%
\bibitem [{\citenamefont {Grusdt}\ and\ \citenamefont
  {Demler}(2015)}]{GrusdtCourse15}%
  \BibitemOpen
  \bibfield  {author} {\bibinfo {author} {\bibfnamefont {F.}~\bibnamefont
  {Grusdt}}\ and\ \bibinfo {author} {\bibfnamefont {E.}~\bibnamefont
  {Demler}},\ }\href {http://arxiv.org/abs/1510.04934} {\bibfield  {journal}
  {\bibinfo  {journal} {arXiv:1510.04934}\ } (\bibinfo {year}
  {2015})}\BibitemShut {NoStop}%
\bibitem [{\citenamefont {Chikkatur}\ \emph {et~al.}(2000)\citenamefont
  {Chikkatur}, \citenamefont {G\"{o}rlitz}, \citenamefont {Stamper-Kurn},
  \citenamefont {Inouye}, \citenamefont {Gupta},\ and\ \citenamefont
  {Ketterle}}]{ChikkaturPRL00}%
  \BibitemOpen
  \bibfield  {author} {\bibinfo {author} {\bibfnamefont {A.~P.}\ \bibnamefont
  {Chikkatur}}, \bibinfo {author} {\bibfnamefont {A.}~\bibnamefont
  {G\"{o}rlitz}}, \bibinfo {author} {\bibfnamefont {D.~M.}\ \bibnamefont
  {Stamper-Kurn}}, \bibinfo {author} {\bibfnamefont {S.}~\bibnamefont
  {Inouye}}, \bibinfo {author} {\bibfnamefont {S.}~\bibnamefont {Gupta}}, \
  and\ \bibinfo {author} {\bibfnamefont {W.}~\bibnamefont {Ketterle}},\ }\href
  {\doibase 10.1103/PhysRevLett.85.483} {\bibfield  {journal} {\bibinfo
  {journal} {Phys. Rev. Lett.}\ }\textbf {\bibinfo {volume} {85}},\ \bibinfo
  {pages} {483} (\bibinfo {year} {2000})}\BibitemShut {NoStop}%
\bibitem [{\citenamefont {Schirotzek}\ \emph {et~al.}(2009)\citenamefont
  {Schirotzek}, \citenamefont {Wu}, \citenamefont {Sommer},\ and\ \citenamefont
  {Zwierlein}}]{SchirotzekPRL09}%
  \BibitemOpen
  \bibfield  {author} {\bibinfo {author} {\bibfnamefont {A.}~\bibnamefont
  {Schirotzek}}, \bibinfo {author} {\bibfnamefont {C.-H.}\ \bibnamefont {Wu}},
  \bibinfo {author} {\bibfnamefont {A.}~\bibnamefont {Sommer}}, \ and\ \bibinfo
  {author} {\bibfnamefont {M.~W.}\ \bibnamefont {Zwierlein}},\ }\href {\doibase
  10.1103/PhysRevLett.102.230402} {\bibfield  {journal} {\bibinfo  {journal}
  {Phys. Rev. Lett.}\ }\textbf {\bibinfo {volume} {102}},\ \bibinfo {pages}
  {230402} (\bibinfo {year} {2009})}\BibitemShut {NoStop}%
\bibitem [{\citenamefont {Palzer}\ \emph {et~al.}(2009)\citenamefont {Palzer},
  \citenamefont {Zipkes}, \citenamefont {Sias},\ and\ \citenamefont
  {K\"{o}hl}}]{PalzerPRL09}%
  \BibitemOpen
  \bibfield  {author} {\bibinfo {author} {\bibfnamefont {S.}~\bibnamefont
  {Palzer}}, \bibinfo {author} {\bibfnamefont {C.}~\bibnamefont {Zipkes}},
  \bibinfo {author} {\bibfnamefont {C.}~\bibnamefont {Sias}}, \ and\ \bibinfo
  {author} {\bibfnamefont {M.}~\bibnamefont {K\"{o}hl}},\ }\href {\doibase
  10.1103/PhysRevLett.103.150601} {\bibfield  {journal} {\bibinfo  {journal}
  {Phys. Rev. Lett.}\ }\textbf {\bibinfo {volume} {103}},\ \bibinfo {pages}
  {150601} (\bibinfo {year} {2009})}\BibitemShut {NoStop}%
\bibitem [{\citenamefont {Kohstall}\ \emph {et~al.}(2012)\citenamefont
  {Kohstall}, \citenamefont {Zaccanti}, \citenamefont {Jag}, \citenamefont
  {Trenkwalder}, \citenamefont {Massignan}, \citenamefont {Bruun},
  \citenamefont {Schreck},\ and\ \citenamefont {Grimm}}]{KohstallNature12}%
  \BibitemOpen
  \bibfield  {author} {\bibinfo {author} {\bibfnamefont {C.}~\bibnamefont
  {Kohstall}}, \bibinfo {author} {\bibfnamefont {M.}~\bibnamefont {Zaccanti}},
  \bibinfo {author} {\bibfnamefont {M.}~\bibnamefont {Jag}}, \bibinfo {author}
  {\bibfnamefont {A.}~\bibnamefont {Trenkwalder}}, \bibinfo {author}
  {\bibfnamefont {P.}~\bibnamefont {Massignan}}, \bibinfo {author}
  {\bibfnamefont {G.~M.}\ \bibnamefont {Bruun}}, \bibinfo {author}
  {\bibfnamefont {F.}~\bibnamefont {Schreck}}, \ and\ \bibinfo {author}
  {\bibfnamefont {R.}~\bibnamefont {Grimm}},\ }\href {\doibase
  10.1038/nature11065} {\bibfield  {journal} {\bibinfo  {journal} {Nature}\
  }\textbf {\bibinfo {volume} {485}},\ \bibinfo {pages} {615} (\bibinfo {year}
  {2012})}\BibitemShut {NoStop}%
\bibitem [{\citenamefont {Koschorreck}\ \emph {et~al.}(2012)\citenamefont
  {Koschorreck}, \citenamefont {Pertot}, \citenamefont {Vogt}, \citenamefont
  {Fr\"ohlich}, \citenamefont {Feld},\ and\ \citenamefont
  {K\"{o}hl}}]{KoschorreckNature12}%
  \BibitemOpen
  \bibfield  {author} {\bibinfo {author} {\bibfnamefont {M.}~\bibnamefont
  {Koschorreck}}, \bibinfo {author} {\bibfnamefont {D.}~\bibnamefont {Pertot}},
  \bibinfo {author} {\bibfnamefont {E.}~\bibnamefont {Vogt}}, \bibinfo {author}
  {\bibfnamefont {B.}~\bibnamefont {Fr\"ohlich}}, \bibinfo {author}
  {\bibfnamefont {M.}~\bibnamefont {Feld}}, \ and\ \bibinfo {author}
  {\bibfnamefont {M.}~\bibnamefont {K\"{o}hl}},\ }\href {\doibase
  10.1038/nature11151} {\bibfield  {journal} {\bibinfo  {journal} {Nature}\
  }\textbf {\bibinfo {volume} {485}},\ \bibinfo {pages} {619} (\bibinfo {year}
  {2012})}\BibitemShut {NoStop}%
\bibitem [{\citenamefont {Spethmann}\ \emph {et~al.}(2012)\citenamefont
  {Spethmann}, \citenamefont {Kindermann}, \citenamefont {John}, \citenamefont
  {Weber}, \citenamefont {Meschede},\ and\ \citenamefont
  {Widera}}]{SpethmannPRL12}%
  \BibitemOpen
  \bibfield  {author} {\bibinfo {author} {\bibfnamefont {N.}~\bibnamefont
  {Spethmann}}, \bibinfo {author} {\bibfnamefont {F.}~\bibnamefont
  {Kindermann}}, \bibinfo {author} {\bibfnamefont {S.}~\bibnamefont {John}},
  \bibinfo {author} {\bibfnamefont {C.}~\bibnamefont {Weber}}, \bibinfo
  {author} {\bibfnamefont {D.}~\bibnamefont {Meschede}}, \ and\ \bibinfo
  {author} {\bibfnamefont {A.}~\bibnamefont {Widera}},\ }\href {\doibase
  10.1103/PhysRevLett.109.235301} {\bibfield  {journal} {\bibinfo  {journal}
  {Phys. Rev. Lett.}\ }\textbf {\bibinfo {volume} {109}},\ \bibinfo {pages}
  {235301} (\bibinfo {year} {2012})}\BibitemShut {NoStop}%
\bibitem [{\citenamefont {Fukuhara}\ \emph {et~al.}(2013)\citenamefont
  {Fukuhara}, \citenamefont {Kantian}, \citenamefont {Endres}, \citenamefont
  {Cheneau}, \citenamefont {Schau\ss}, \citenamefont {Hild}, \citenamefont
  {Bellem}, \citenamefont {Schollw\"{o}ck}, \citenamefont {Giamarchi},
  \citenamefont {Gross}, \citenamefont {Bloch},\ and\ \citenamefont
  {Kuhr}}]{FukuharaNatPhys13}%
  \BibitemOpen
  \bibfield  {author} {\bibinfo {author} {\bibfnamefont {T.}~\bibnamefont
  {Fukuhara}}, \bibinfo {author} {\bibfnamefont {A.}~\bibnamefont {Kantian}},
  \bibinfo {author} {\bibfnamefont {M.}~\bibnamefont {Endres}}, \bibinfo
  {author} {\bibfnamefont {M.}~\bibnamefont {Cheneau}}, \bibinfo {author}
  {\bibfnamefont {P.}~\bibnamefont {Schau\ss}}, \bibinfo {author}
  {\bibfnamefont {S.}~\bibnamefont {Hild}}, \bibinfo {author} {\bibfnamefont
  {D.}~\bibnamefont {Bellem}}, \bibinfo {author} {\bibfnamefont
  {U.}~\bibnamefont {Schollw\"{o}ck}}, \bibinfo {author} {\bibfnamefont
  {T.}~\bibnamefont {Giamarchi}}, \bibinfo {author} {\bibfnamefont
  {C.}~\bibnamefont {Gross}}, \bibinfo {author} {\bibfnamefont
  {I.}~\bibnamefont {Bloch}}, \ and\ \bibinfo {author} {\bibfnamefont
  {S.}~\bibnamefont {Kuhr}},\ }\href {\doibase 10.1038/nphys2561} {\bibfield
  {journal} {\bibinfo  {journal} {Nat. Phys.}\ }\textbf {\bibinfo {volume}
  {9}},\ \bibinfo {pages} {235} (\bibinfo {year} {2013})}\BibitemShut {NoStop}%
\bibitem [{\citenamefont {Scelle}\ \emph {et~al.}(2013)\citenamefont {Scelle},
  \citenamefont {Rentrop}, \citenamefont {Trautmann}, \citenamefont
  {Schuster},\ and\ \citenamefont {Oberthaler}}]{ScellePRL13}%
  \BibitemOpen
  \bibfield  {author} {\bibinfo {author} {\bibfnamefont {R.}~\bibnamefont
  {Scelle}}, \bibinfo {author} {\bibfnamefont {T.}~\bibnamefont {Rentrop}},
  \bibinfo {author} {\bibfnamefont {A.}~\bibnamefont {Trautmann}}, \bibinfo
  {author} {\bibfnamefont {T.}~\bibnamefont {Schuster}}, \ and\ \bibinfo
  {author} {\bibfnamefont {M.~K.}\ \bibnamefont {Oberthaler}},\ }\href
  {\doibase 10.1103/PhysRevLett.111.070401} {\bibfield  {journal} {\bibinfo
  {journal} {Phys. Rev. Lett.}\ }\textbf {\bibinfo {volume} {111}},\ \bibinfo
  {pages} {070401} (\bibinfo {year} {2013})}\BibitemShut {NoStop}%
\bibitem [{\citenamefont {Cetina}\ \emph {et~al.}(2015)\citenamefont {Cetina},
  \citenamefont {Jag}, \citenamefont {Lous}, \citenamefont {Walraven},
  \citenamefont {Grimm}, \citenamefont {Christensen},\ and\ \citenamefont
  {Bruun}}]{Cetina15}%
  \BibitemOpen
  \bibfield  {author} {\bibinfo {author} {\bibfnamefont {M.}~\bibnamefont
  {Cetina}}, \bibinfo {author} {\bibfnamefont {M.}~\bibnamefont {Jag}},
  \bibinfo {author} {\bibfnamefont {R.~S.}\ \bibnamefont {Lous}}, \bibinfo
  {author} {\bibfnamefont {J.~T.~M.}\ \bibnamefont {Walraven}}, \bibinfo
  {author} {\bibfnamefont {R.}~\bibnamefont {Grimm}}, \bibinfo {author}
  {\bibfnamefont {R.~S.}\ \bibnamefont {Christensen}}, \ and\ \bibinfo {author}
  {\bibfnamefont {G.~M.}\ \bibnamefont {Bruun}},\ }\href {\doibase
  10.1103/PhysRevLett.115.135302} {\bibfield  {journal} {\bibinfo  {journal}
  {Phys. Rev. Lett.}\ }\textbf {\bibinfo {volume} {115}},\ \bibinfo {pages}
  {135302} (\bibinfo {year} {2015})}\BibitemShut {NoStop}%
\bibitem [{\citenamefont {Massignan}\ \emph {et~al.}(2014)\citenamefont
  {Massignan}, \citenamefont {Zaccanti},\ and\ \citenamefont
  {Bruun}}]{MassignanRPP14}%
  \BibitemOpen
  \bibfield  {author} {\bibinfo {author} {\bibfnamefont {P.}~\bibnamefont
  {Massignan}}, \bibinfo {author} {\bibfnamefont {M.}~\bibnamefont {Zaccanti}},
  \ and\ \bibinfo {author} {\bibfnamefont {G.~M.}\ \bibnamefont {Bruun}},\
  }\href {\doibase 10.1088/0034-4885/77/3/034401} {\bibfield  {journal}
  {\bibinfo  {journal} {Rep. Prog. Phys.}\ }\textbf {\bibinfo {volume} {77}},\
  \bibinfo {pages} {034401} (\bibinfo {year} {2014})}\BibitemShut {NoStop}%
\bibitem [{\citenamefont {{J{\o}rgensen}}\ \emph {et~al.}()\citenamefont
  {{J{\o}rgensen}}, \citenamefont {{Wacker}}, \citenamefont {{Skalmstang}},
  \citenamefont {{Parish}}, \citenamefont {{Levinsen}}, \citenamefont
  {{Christensen}}, \citenamefont {{Bruun}},\ and\ \citenamefont
  {{Arlt}}}]{Jorgensen2016}%
  \BibitemOpen
  \bibfield  {author} {\bibinfo {author} {\bibfnamefont {N.~B.}\ \bibnamefont
  {{J{\o}rgensen}}}, \bibinfo {author} {\bibfnamefont {L.}~\bibnamefont
  {{Wacker}}}, \bibinfo {author} {\bibfnamefont {K.~T.}\ \bibnamefont
  {{Skalmstang}}}, \bibinfo {author} {\bibfnamefont {M.~M.}\ \bibnamefont
  {{Parish}}}, \bibinfo {author} {\bibfnamefont {J.}~\bibnamefont
  {{Levinsen}}}, \bibinfo {author} {\bibfnamefont {R.~S.}\ \bibnamefont
  {{Christensen}}}, \bibinfo {author} {\bibfnamefont {G.~M.}\ \bibnamefont
  {{Bruun}}}, \ and\ \bibinfo {author} {\bibfnamefont {J.~J.}\ \bibnamefont
  {{Arlt}}},\ }\href {http://arxiv.org/abs/1604.07883} {\bibinfo  {journal}
  {arXiv:1604.07883}\ }\BibitemShut {NoStop}%
\bibitem [{\citenamefont {Hu}\ \emph {et~al.}()\citenamefont {Hu},
  \citenamefont {de~Graaff}, \citenamefont {Kedar}, \citenamefont {Corson},
  \citenamefont {Cornell},\ and\ \citenamefont {Jin}}]{Hu16}%
  \BibitemOpen
\bibfield  {journal} {  }\bibfield  {author} {\bibinfo {author} {\bibfnamefont
  {M.-G.}\ \bibnamefont {Hu}}, \bibinfo {author} {\bibfnamefont {M.~J.~V.}\
  \bibnamefont {de~Graaff}}, \bibinfo {author} {\bibfnamefont {D.}~\bibnamefont
  {Kedar}}, \bibinfo {author} {\bibfnamefont {J.~P.}\ \bibnamefont {Corson}},
  \bibinfo {author} {\bibfnamefont {E.~A.}\ \bibnamefont {Cornell}}, \ and\
  \bibinfo {author} {\bibfnamefont {D.~S.}\ \bibnamefont {Jin}},\ }\href
  {http://arxiv.org/abs/1605.00729} {\bibinfo  {journal} {arXiv:1605.00729}\
  }\BibitemShut {NoStop}%
\bibitem [{\citenamefont {Cetina}\ \emph {et~al.}()\citenamefont {Cetina},
  \citenamefont {Jag}, \citenamefont {Lous}, \citenamefont {Fritsche},
  \citenamefont {Walraven}, \citenamefont {Grimm}, \citenamefont {Levinsen},
  \citenamefont {Parish}, \citenamefont {Schmidt}, \citenamefont {Knap} \emph
  {et~al.}}]{Cetina2016}%
  \BibitemOpen
\bibfield  {journal} {  }\bibfield  {author} {\bibinfo {author} {\bibfnamefont
  {M.}~\bibnamefont {Cetina}}, \bibinfo {author} {\bibfnamefont
  {M.}~\bibnamefont {Jag}}, \bibinfo {author} {\bibfnamefont {R.}~\bibnamefont
  {Lous}}, \bibinfo {author} {\bibfnamefont {I.}~\bibnamefont {Fritsche}},
  \bibinfo {author} {\bibfnamefont {J.}~\bibnamefont {Walraven}}, \bibinfo
  {author} {\bibfnamefont {R.}~\bibnamefont {Grimm}}, \bibinfo {author}
  {\bibfnamefont {J.}~\bibnamefont {Levinsen}}, \bibinfo {author}
  {\bibfnamefont {M.}~\bibnamefont {Parish}}, \bibinfo {author} {\bibfnamefont
  {R.}~\bibnamefont {Schmidt}}, \bibinfo {author} {\bibfnamefont
  {M.}~\bibnamefont {Knap}},  \emph {et~al.},\ }\href
  {http://arxiv.org/abs/1604.07423} {\bibinfo  {journal} {arXiv:1604.07423}\
  }\BibitemShut {NoStop}%
\bibitem [{\citenamefont {Gonz\'alez-S\'anchez}\ \emph
  {et~al.}(2015{\natexlab{a}})\citenamefont {Gonz\'alez-S\'anchez},
  \citenamefont {Carelli}, \citenamefont {Gianturco},\ and\ \citenamefont
  {Wester}}]{GonzalezCP15}%
  \BibitemOpen
\bibfield  {journal} {  }\bibfield  {author} {\bibinfo {author} {\bibfnamefont
  {L.}~\bibnamefont {Gonz\'alez-S\'anchez}}, \bibinfo {author} {\bibfnamefont
  {F.}~\bibnamefont {Carelli}}, \bibinfo {author} {\bibfnamefont
  {F.}~\bibnamefont {Gianturco}}, \ and\ \bibinfo {author} {\bibfnamefont
  {R.}~\bibnamefont {Wester}},\ }\href {\doibase
  10.1016/j.chemphys.2015.05.027} {\bibfield  {journal} {\bibinfo  {journal}
  {Chem. Phys.}\ }\textbf {\bibinfo {volume} {462}},\ \bibinfo {pages} {111}
  (\bibinfo {year} {2015}{\natexlab{a}})}\BibitemShut {NoStop}%
\bibitem [{\citenamefont {Gonz\'alez-S\'anchez}\ \emph
  {et~al.}(2015{\natexlab{b}})\citenamefont {Gonz\'alez-S\'anchez},
  \citenamefont {Gianturco}, \citenamefont {Carelli},\ and\ \citenamefont
  {Wester}}]{GonzalezNJP15}%
  \BibitemOpen
  \bibfield  {author} {\bibinfo {author} {\bibfnamefont {L.}~\bibnamefont
  {Gonz\'alez-S\'anchez}}, \bibinfo {author} {\bibfnamefont {F.~A.}\
  \bibnamefont {Gianturco}}, \bibinfo {author} {\bibfnamefont {F.}~\bibnamefont
  {Carelli}}, \ and\ \bibinfo {author} {\bibfnamefont {R.}~\bibnamefont
  {Wester}},\ }\href {\doibase 10.1088/1367-2630/17/12/123003} {\bibfield
  {journal} {\bibinfo  {journal} {New J. Phys.}\ }\textbf {\bibinfo {volume}
  {17}},\ \bibinfo {pages} {123003} (\bibinfo {year}
  {2015}{\natexlab{b}})}\BibitemShut {NoStop}%
\bibitem [{\citenamefont {Mikosch}\ \emph {et~al.}(2010)\citenamefont
  {Mikosch}, \citenamefont {Weidem\"{u}ller},\ and\ \citenamefont
  {Wester}}]{MikoschIRPC}%
  \BibitemOpen
  \bibfield  {author} {\bibinfo {author} {\bibfnamefont {J.}~\bibnamefont
  {Mikosch}}, \bibinfo {author} {\bibfnamefont {M.}~\bibnamefont
  {Weidem\"{u}ller}}, \ and\ \bibinfo {author} {\bibfnamefont {R.}~\bibnamefont
  {Wester}},\ }\href {\doibase 10.1080/0144235X.2010.519504} {\bibfield
  {journal} {\bibinfo  {journal} {Int. Rev. Phys. Chem.}\ }\textbf {\bibinfo
  {volume} {29}},\ \bibinfo {pages} {589} (\bibinfo {year} {2010})}\BibitemShut
  {NoStop}%
\bibitem [{\citenamefont {Wester}(2011)}]{WesterHandSpec11}%
  \BibitemOpen
  \bibfield  {author} {\bibinfo {author} {\bibfnamefont {R.}~\bibnamefont
  {Wester}},\ }in\ \href@noop {} {\emph {\bibinfo {booktitle} {Handbook of
  High-resolution Spectroscopy}}}\ (\bibinfo  {publisher} {Wiley Online
  Library},\ \bibinfo {year} {2011})\BibitemShut {NoStop}%
\bibitem [{\citenamefont {Bradforth}\ \emph {et~al.}(1993)\citenamefont
  {Bradforth}, \citenamefont {Kim}, \citenamefont {Arnold},\ and\ \citenamefont
  {Neumark}}]{BradforthJCP93}%
  \BibitemOpen
  \bibfield  {author} {\bibinfo {author} {\bibfnamefont {S.~E.}\ \bibnamefont
  {Bradforth}}, \bibinfo {author} {\bibfnamefont {E.~H.}\ \bibnamefont {Kim}},
  \bibinfo {author} {\bibfnamefont {D.~W.}\ \bibnamefont {Arnold}}, \ and\
  \bibinfo {author} {\bibfnamefont {D.~M.}\ \bibnamefont {Neumark}},\ }\href
  {\doibase 10.1063/1.464244} {\bibfield  {journal} {\bibinfo  {journal} {J.
  Chem. Phys.}\ }\textbf {\bibinfo {volume} {98}},\ \bibinfo {pages} {800}
  (\bibinfo {year} {1993})}\BibitemShut {NoStop}%
\bibitem [{\citenamefont {Bartlett}\ and\ \citenamefont
  {Musial}(2007)}]{MusialRMP07}%
  \BibitemOpen
  \bibfield  {author} {\bibinfo {author} {\bibfnamefont {R.~J.}\ \bibnamefont
  {Bartlett}}\ and\ \bibinfo {author} {\bibfnamefont {M.}~\bibnamefont
  {Musial}},\ }\href {\doibase 10.1103/RevModPhys.79.291} {\bibfield  {journal}
  {\bibinfo  {journal} {Rev. Mod. Phys.}\ }\textbf {\bibinfo {volume} {79}},\
  \bibinfo {pages} {291} (\bibinfo {year} {2007})}\BibitemShut {NoStop}%
\bibitem [{\citenamefont {Boys}\ and\ \citenamefont
  {Bernardi}(1970)}]{BoysMP70}%
  \BibitemOpen
  \bibfield  {author} {\bibinfo {author} {\bibfnamefont {S.}~\bibnamefont
  {Boys}}\ and\ \bibinfo {author} {\bibfnamefont {F.}~\bibnamefont
  {Bernardi}},\ }\href {\doibase 10.1080/00268977000101561} {\bibfield
  {journal} {\bibinfo  {journal} {Mol. Phys.}\ }\textbf {\bibinfo {volume}
  {19}},\ \bibinfo {pages} {553} (\bibinfo {year} {1970})}\BibitemShut
  {NoStop}%
\bibitem [{\citenamefont {Dunning}(1989)}]{DunningJCP89}%
  \BibitemOpen
  \bibfield  {author} {\bibinfo {author} {\bibfnamefont {J.~T.~H.}\
  \bibnamefont {Dunning}},\ }\href {\doibase 10.1063/1.456153} {\bibfield
  {journal} {\bibinfo  {journal} {J. Chem. Phys.}\ }\textbf {\bibinfo {volume}
  {90}},\ \bibinfo {pages} {1007} (\bibinfo {year} {1989})}\BibitemShut
  {NoStop}%
\bibitem [{\citenamefont {Dolg}\ and\ \citenamefont {Cao}(2012)}]{DolgCR12}%
  \BibitemOpen
  \bibfield  {author} {\bibinfo {author} {\bibfnamefont {M.}~\bibnamefont
  {Dolg}}\ and\ \bibinfo {author} {\bibfnamefont {X.}~\bibnamefont {Cao}},\
  }\href {\doibase 10.1021/cr2001383} {\bibfield  {journal} {\bibinfo
  {journal} {Chem. Rev.}\ }\textbf {\bibinfo {volume} {112}},\ \bibinfo {pages}
  {403} (\bibinfo {year} {2012})}\BibitemShut {NoStop}%
\bibitem [{\citenamefont {Lim}\ \emph {et~al.}(2005)\citenamefont {Lim},
  \citenamefont {Schwerdtfeger}, \citenamefont {Metz},\ and\ \citenamefont
  {Stoll}}]{LimJCP05}%
  \BibitemOpen
  \bibfield  {author} {\bibinfo {author} {\bibfnamefont {I.~S.}\ \bibnamefont
  {Lim}}, \bibinfo {author} {\bibfnamefont {P.}~\bibnamefont {Schwerdtfeger}},
  \bibinfo {author} {\bibfnamefont {B.}~\bibnamefont {Metz}}, \ and\ \bibinfo
  {author} {\bibfnamefont {H.}~\bibnamefont {Stoll}},\ }\href {\doibase
  10.1063/1.1856451} {\bibfield  {journal} {\bibinfo  {journal} {J. Chem.
  Phys.}\ }\textbf {\bibinfo {volume} {122}},\ \bibinfo {pages} {104103}
  (\bibinfo {year} {2005})}\BibitemShut {NoStop}%
\bibitem [{\citenamefont {Lim}\ \emph {et~al.}(2006)\citenamefont {Lim},
  \citenamefont {Stoll},\ and\ \citenamefont {Schwerdtfeger}}]{LimJCP06}%
  \BibitemOpen
  \bibfield  {author} {\bibinfo {author} {\bibfnamefont {I.~S.}\ \bibnamefont
  {Lim}}, \bibinfo {author} {\bibfnamefont {H.}~\bibnamefont {Stoll}}, \ and\
  \bibinfo {author} {\bibfnamefont {P.}~\bibnamefont {Schwerdtfeger}},\ }\href
  {\doibase 10.1063/1.2148945} {\bibfield  {journal} {\bibinfo  {journal} {J.
  Chem. Phys.}\ }\textbf {\bibinfo {volume} {124}},\ \bibinfo {pages} {034107}
  (\bibinfo {year} {2006})}\BibitemShut {NoStop}%
\bibitem [{\citenamefont {Tomza}\ \emph {et~al.}(2013)\citenamefont {Tomza},
  \citenamefont {Skomorowski}, \citenamefont {Musial}, \citenamefont {Ferez},
  \citenamefont {Koch},\ and\ \citenamefont {Moszynski}}]{TomzaMP13}%
  \BibitemOpen
  \bibfield  {author} {\bibinfo {author} {\bibfnamefont {M.}~\bibnamefont
  {Tomza}}, \bibinfo {author} {\bibfnamefont {W.}~\bibnamefont {Skomorowski}},
  \bibinfo {author} {\bibfnamefont {M.}~\bibnamefont {Musial}}, \bibinfo
  {author} {\bibfnamefont {R.~G.}\ \bibnamefont {Ferez}}, \bibinfo {author}
  {\bibfnamefont {C.~P.}\ \bibnamefont {Koch}}, \ and\ \bibinfo {author}
  {\bibfnamefont {R.}~\bibnamefont {Moszynski}},\ }\href {\doibase
  10.1080/00268976.2013.793835} {\bibfield  {journal} {\bibinfo  {journal}
  {Mol. Phys.}\ }\textbf {\bibinfo {volume} {111}},\ \bibinfo {pages} {1781}
  (\bibinfo {year} {2013})}\BibitemShut {NoStop}%
\bibitem [{\citenamefont {Tomza}\ \emph {et~al.}(2011)\citenamefont {Tomza},
  \citenamefont {Pawlowski}, \citenamefont {Jeziorska}, \citenamefont {Koch},\
  and\ \citenamefont {Moszynski}}]{TomzaPCCP11}%
  \BibitemOpen
  \bibfield  {author} {\bibinfo {author} {\bibfnamefont {M.}~\bibnamefont
  {Tomza}}, \bibinfo {author} {\bibfnamefont {F.}~\bibnamefont {Pawlowski}},
  \bibinfo {author} {\bibfnamefont {M.}~\bibnamefont {Jeziorska}}, \bibinfo
  {author} {\bibfnamefont {C.~P.}\ \bibnamefont {Koch}}, \ and\ \bibinfo
  {author} {\bibfnamefont {R.}~\bibnamefont {Moszynski}},\ }\href {\doibase
  10.1039/C1CP21196J} {\bibfield  {journal} {\bibinfo  {journal} {Phys. Chem.
  Chem. Phys.}\ }\textbf {\bibinfo {volume} {13}},\ \bibinfo {pages} {18893}
  (\bibinfo {year} {2011})}\BibitemShut {NoStop}%
\bibitem [{\citenamefont {Werner}\ \emph {et~al.}(2012)\citenamefont {Werner},
  \citenamefont {Knowles}, \citenamefont {Lindh}, \citenamefont {Sch\"utz},
  \citenamefont {Celani}, \citenamefont {Korona}, \citenamefont {Rauhut},
  \citenamefont {D.Amos}, \citenamefont {Bernhardsson}, \citenamefont {Berning}
  \emph {et~al.}}]{Molpro}%
  \BibitemOpen
  \bibfield  {author} {\bibinfo {author} {\bibfnamefont {H.-J.}\ \bibnamefont
  {Werner}}, \bibinfo {author} {\bibfnamefont {P.~J.}\ \bibnamefont {Knowles}},
  \bibinfo {author} {\bibfnamefont {F.~R. M.~R.}\ \bibnamefont {Lindh}},
  \bibinfo {author} {\bibfnamefont {M.}~\bibnamefont {Sch\"utz}}, \bibinfo
  {author} {\bibfnamefont {P.}~\bibnamefont {Celani}}, \bibinfo {author}
  {\bibfnamefont {T.}~\bibnamefont {Korona}}, \bibinfo {author} {\bibfnamefont
  {G.}~\bibnamefont {Rauhut}}, \bibinfo {author} {\bibfnamefont
  {R.}~\bibnamefont {D.Amos}}, \bibinfo {author} {\bibfnamefont
  {A.}~\bibnamefont {Bernhardsson}}, \bibinfo {author} {\bibfnamefont
  {A.}~\bibnamefont {Berning}},  \emph {et~al.},\ }\href
  {http://www.molpro.net} {\enquote {\bibinfo {title} {Molpro, version 2012.1,
  a package of ab initio programs},}\ } (\bibinfo {year} {2012})\BibitemShut
  {NoStop}%
\bibitem [{\citenamefont {Jeziorski}\ \emph {et~al.}(1994)\citenamefont
  {Jeziorski}, \citenamefont {Moszynski},\ and\ \citenamefont
  {Szalewicz}}]{JeziorskiCR94}%
  \BibitemOpen
  \bibfield  {author} {\bibinfo {author} {\bibfnamefont {B.}~\bibnamefont
  {Jeziorski}}, \bibinfo {author} {\bibfnamefont {R.}~\bibnamefont
  {Moszynski}}, \ and\ \bibinfo {author} {\bibfnamefont {K.}~\bibnamefont
  {Szalewicz}},\ }\href {\doibase 10.1021/cr00031a008} {\bibfield  {journal}
  {\bibinfo  {journal} {Chem. Rev.}\ }\textbf {\bibinfo {volume} {94}},\
  \bibinfo {pages} {1887} (\bibinfo {year} {1994})}\BibitemShut {NoStop}%
\bibitem [{\citenamefont {Derevianko}\ \emph {et~al.}(2010)\citenamefont
  {Derevianko}, \citenamefont {Porsev},\ and\ \citenamefont
  {Babb}}]{DerevienkoADNDT10}%
  \BibitemOpen
  \bibfield  {author} {\bibinfo {author} {\bibfnamefont {A.}~\bibnamefont
  {Derevianko}}, \bibinfo {author} {\bibfnamefont {S.~G.}\ \bibnamefont
  {Porsev}}, \ and\ \bibinfo {author} {\bibfnamefont {J.~F.}\ \bibnamefont
  {Babb}},\ }\href {\doibase 10.1016/j.adt.2009.12.002} {\bibfield  {journal}
  {\bibinfo  {journal} {At. Data Nucl. Data Tables}\ }\textbf {\bibinfo
  {volume} {96}},\ \bibinfo {pages} {323} (\bibinfo {year} {2010})}\BibitemShut
  {NoStop}%
\bibitem [{\citenamefont {Moszynski}\ \emph {et~al.}(2005)\citenamefont
  {Moszynski}, \citenamefont {Zuchowski},\ and\ \citenamefont
  {Jeziorsk}}]{MoszynskiCCCC05}%
  \BibitemOpen
  \bibfield  {author} {\bibinfo {author} {\bibfnamefont {R.}~\bibnamefont
  {Moszynski}}, \bibinfo {author} {\bibfnamefont {P.}~\bibnamefont
  {Zuchowski}}, \ and\ \bibinfo {author} {\bibfnamefont {B.}~\bibnamefont
  {Jeziorsk}},\ }\href {\doibase 10.1135/cccc20051109} {\bibfield  {journal}
  {\bibinfo  {journal} {Collect. Czech. Chem. Commun.}\ }\textbf {\bibinfo
  {volume} {70}},\ \bibinfo {pages} {1109} (\bibinfo {year}
  {2005})}\BibitemShut {NoStop}%
\bibitem [{\citenamefont {Korona}\ \emph {et~al.}(2006)\citenamefont {Korona},
  \citenamefont {Przybytek},\ and\ \citenamefont {Jeziorski}}]{KoronaMP06}%
  \BibitemOpen
  \bibfield  {author} {\bibinfo {author} {\bibfnamefont {T.}~\bibnamefont
  {Korona}}, \bibinfo {author} {\bibfnamefont {M.}~\bibnamefont {Przybytek}}, \
  and\ \bibinfo {author} {\bibfnamefont {B.}~\bibnamefont {Jeziorski}},\ }\href
  {\doibase 10.1080/00268970600673975} {\bibfield  {journal} {\bibinfo
  {journal} {Mol. Phys.}\ }\textbf {\bibinfo {volume} {104}},\ \bibinfo {pages}
  {2303} (\bibinfo {year} {2006})}\BibitemShut {NoStop}%
\bibitem [{\citenamefont {Rath}\ and\ \citenamefont
  {Schmidt}(2013)}]{Rath2013}%
  \BibitemOpen
  \bibfield  {author} {\bibinfo {author} {\bibfnamefont {S.~P.}\ \bibnamefont
  {Rath}}\ and\ \bibinfo {author} {\bibfnamefont {R.}~\bibnamefont {Schmidt}},\
  }\href {\doibase 10.1103/PhysRevA.88.053632} {\bibfield  {journal} {\bibinfo
  {journal} {Phys. Rev. A}\ }\textbf {\bibinfo {volume} {88}},\ \bibinfo
  {pages} {053632} (\bibinfo {year} {2013})}\BibitemShut {NoStop}%
\bibitem [{\citenamefont {Girardeau}(1961)}]{GirardeauPF61}%
  \BibitemOpen
  \bibfield  {author} {\bibinfo {author} {\bibfnamefont {M.}~\bibnamefont
  {Girardeau}},\ }\href {\doibase 10.1063/1.1706323} {\bibfield  {journal}
  {\bibinfo  {journal} {Phys. Fluids}\ }\textbf {\bibinfo {volume} {4}},\
  \bibinfo {pages} {279} (\bibinfo {year} {1961})}\BibitemShut {NoStop}%
\bibitem [{\citenamefont {Shchadilova}\ \emph {et~al.}(2016)\citenamefont
  {Shchadilova}, \citenamefont {Schmidt}, \citenamefont {Grusdt},\ and\
  \citenamefont {Demler}}]{Shchadilova2016}%
  \BibitemOpen
  \bibfield  {author} {\bibinfo {author} {\bibfnamefont {Y.~E.}\ \bibnamefont
  {Shchadilova}}, \bibinfo {author} {\bibfnamefont {R.}~\bibnamefont
  {Schmidt}}, \bibinfo {author} {\bibfnamefont {F.}~\bibnamefont {Grusdt}}, \
  and\ \bibinfo {author} {\bibfnamefont {E.}~\bibnamefont {Demler}},\ }\href
  {https://arxiv.org/abs/1604.06469} {\bibfield  {journal} {\bibinfo  {journal}
  {arXiv:1604.06469}\ } (\bibinfo {year} {2016})}\BibitemShut {NoStop}%
\bibitem [{\citenamefont {Friedrich}(2013)}]{friedrich2013scattering}%
  \BibitemOpen
  \bibfield  {author} {\bibinfo {author} {\bibfnamefont {H.}~\bibnamefont
  {Friedrich}},\ }\href@noop {} {\emph {\bibinfo {title} {Scattering
  theory}}},\ Vol.\ \bibinfo {volume} {872}\ (\bibinfo  {publisher}
  {Springer},\ \bibinfo {year} {2013})\BibitemShut {NoStop}%
\bibitem [{\citenamefont {Gottlieb}\ \emph {et~al.}(2007)\citenamefont
  {Gottlieb}, \citenamefont {Br\"unken}, \citenamefont {McCarthy},\ and\
  \citenamefont {Thaddeus}}]{Gottlieb07}%
  \BibitemOpen
  \bibfield  {author} {\bibinfo {author} {\bibfnamefont {C.~A.}\ \bibnamefont
  {Gottlieb}}, \bibinfo {author} {\bibfnamefont {S.}~\bibnamefont {Br\"unken}},
  \bibinfo {author} {\bibfnamefont {M.~C.}\ \bibnamefont {McCarthy}}, \ and\
  \bibinfo {author} {\bibfnamefont {P.}~\bibnamefont {Thaddeus}},\ }\href
  {\doibase 10.1063/1.2737442} {\bibfield  {journal} {\bibinfo  {journal} {J.
  Chem. Phys.}\ }\textbf {\bibinfo {volume} {126}},\ \bibinfo {pages} {191101}
  (\bibinfo {year} {2007})}\BibitemShut {NoStop}%
\bibitem [{\citenamefont {van Kempen}\ \emph {et~al.}(2002)\citenamefont {van
  Kempen}, \citenamefont {Kokkelmans}, \citenamefont {Heinzen},\ and\
  \citenamefont {Verhaar}}]{KempenPRL2002}%
  \BibitemOpen
  \bibfield  {author} {\bibinfo {author} {\bibfnamefont {E.~G.~M.}\
  \bibnamefont {van Kempen}}, \bibinfo {author} {\bibfnamefont {S.~J. J.
  M.~F.}\ \bibnamefont {Kokkelmans}}, \bibinfo {author} {\bibfnamefont {D.~J.}\
  \bibnamefont {Heinzen}}, \ and\ \bibinfo {author} {\bibfnamefont {B.~J.}\
  \bibnamefont {Verhaar}},\ }\href {\doibase 10.1103/PhysRevLett.88.093201}
  {\bibfield  {journal} {\bibinfo  {journal} {Phys. Rev. Lett.}\ }\textbf
  {\bibinfo {volume} {88}},\ \bibinfo {pages} {093201} (\bibinfo {year}
  {2002})}\BibitemShut {NoStop}%
\bibitem [{\citenamefont {de~Escobar}\ \emph {et~al.}(2009)\citenamefont
  {de~Escobar}, \citenamefont {Mickelson}, \citenamefont {Yan}, \citenamefont
  {DeSalvo}, \citenamefont {Nagel},\ and\ \citenamefont
  {Killian}}]{KillianPRL2009}%
  \BibitemOpen
  \bibfield  {author} {\bibinfo {author} {\bibfnamefont {Y.~N.~M.}\
  \bibnamefont {de~Escobar}}, \bibinfo {author} {\bibfnamefont {P.~G.}\
  \bibnamefont {Mickelson}}, \bibinfo {author} {\bibfnamefont {M.}~\bibnamefont
  {Yan}}, \bibinfo {author} {\bibfnamefont {B.~J.}\ \bibnamefont {DeSalvo}},
  \bibinfo {author} {\bibfnamefont {S.~B.}\ \bibnamefont {Nagel}}, \ and\
  \bibinfo {author} {\bibfnamefont {T.~C.}\ \bibnamefont {Killian}},\ }\href
  {\doibase 10.1103/PhysRevLett.103.200402} {\bibfield  {journal} {\bibinfo
  {journal} {Phys. Rev. Lett.}\ }\textbf {\bibinfo {volume} {103}},\ \bibinfo
  {pages} {200402} (\bibinfo {year} {2009})}\BibitemShut {NoStop}%
\bibitem [{\citenamefont {Rentrop}\ \emph {et~al.}(2016)\citenamefont
  {Rentrop}, \citenamefont {Trautmann}, \citenamefont {Olivares}, \citenamefont
  {Jendrzejewski}, \citenamefont {Komnik},\ and\ \citenamefont
  {Oberthaler}}]{oberthaler}%
  \BibitemOpen
  \bibfield  {author} {\bibinfo {author} {\bibfnamefont {T.}~\bibnamefont
  {Rentrop}}, \bibinfo {author} {\bibfnamefont {A.}~\bibnamefont {Trautmann}},
  \bibinfo {author} {\bibfnamefont {F.}~\bibnamefont {Olivares}}, \bibinfo
  {author} {\bibfnamefont {F.}~\bibnamefont {Jendrzejewski}}, \bibinfo {author}
  {\bibfnamefont {A.}~\bibnamefont {Komnik}}, \ and\ \bibinfo {author}
  {\bibfnamefont {M.}~\bibnamefont {Oberthaler}},\ }\href
  {https://arxiv.org/abs/1605.01874} {\bibfield  {journal} {\bibinfo  {journal}
  {arXiv:1605.01874}\ } (\bibinfo {year} {2016})}\BibitemShut {NoStop}%
\bibitem [{\citenamefont {Krych}\ and\ \citenamefont
  {Idziaszek}(2015)}]{KrychPRA15}%
  \BibitemOpen
  \bibfield  {author} {\bibinfo {author} {\bibfnamefont {M.}~\bibnamefont
  {Krych}}\ and\ \bibinfo {author} {\bibfnamefont {Z.}~\bibnamefont
  {Idziaszek}},\ }\href@noop {} {\bibfield  {journal} {\bibinfo  {journal}
  {Phys. Rev. A}\ }\textbf {\bibinfo {volume} {91}},\ \bibinfo {pages} {023430}
  (\bibinfo {year} {2015})}\BibitemShut {NoStop}%
\bibitem [{\citenamefont {Midya}\ \emph {et~al.}()\citenamefont {Midya} \emph
  {et~al.}}]{MidyaPreparation}%
  \BibitemOpen
  \bibfield  {author} {\bibinfo {author} {\bibfnamefont {B.}~\bibnamefont
  {Midya}} \emph {et~al.},\ }\href@noop {} {\bibinfo  {journal}
  {\textit{(unpublished)}}\ }\BibitemShut {NoStop}%
\bibitem [{\citenamefont {Prokof'ev}\ and\ \citenamefont
  {Svistunov}(2008)}]{prok2008}%
  \BibitemOpen
\bibfield  {journal} {  }\bibfield  {author} {\bibinfo {author} {\bibfnamefont
  {N.}~\bibnamefont {Prokof'ev}}\ and\ \bibinfo {author} {\bibfnamefont
  {B.}~\bibnamefont {Svistunov}},\ }\href {\doibase 10.1103/PhysRevB.77.020408}
  {\bibfield  {journal} {\bibinfo  {journal} {Phys. Rev. B}\ }\textbf {\bibinfo
  {volume} {77}},\ \bibinfo {pages} {020408} (\bibinfo {year}
  {2008})}\BibitemShut {NoStop}%
\bibitem [{\citenamefont {Schmidt}\ and\ \citenamefont
  {Enss}(2011)}]{schmidt_excitation_2011}%
  \BibitemOpen
  \bibfield  {author} {\bibinfo {author} {\bibfnamefont {R.}~\bibnamefont
  {Schmidt}}\ and\ \bibinfo {author} {\bibfnamefont {T.}~\bibnamefont {Enss}},\
  }\href {\doibase 10.1103/PhysRevA.83.063620} {\bibfield  {journal} {\bibinfo
  {journal} {Phys. Rev. A}\ }\textbf {\bibinfo {volume} {83}},\ \bibinfo
  {pages} {063620} (\bibinfo {year} {2011})}\BibitemShut {NoStop}%
\bibitem [{\citenamefont {Anderson}(1967)}]{anderson1967}%
  \BibitemOpen
  \bibfield  {author} {\bibinfo {author} {\bibfnamefont {P.~W.}\ \bibnamefont
  {Anderson}},\ }\href {\doibase 10.1103/PhysRevLett.18.1049} {\bibfield
  {journal} {\bibinfo  {journal} {Phys. Rev. Lett.}\ }\textbf {\bibinfo
  {volume} {18}},\ \bibinfo {pages} {1049} (\bibinfo {year}
  {1967})}\BibitemShut {NoStop}%
\bibitem [{\citenamefont {Lahaye}\ \emph {et~al.}(2009)\citenamefont {Lahaye},
  \citenamefont {Menotti}, \citenamefont {Santos}, \citenamefont {Lewenstein},\
  and\ \citenamefont {Pfau}}]{LahayePfauRPP2009}%
  \BibitemOpen
  \bibfield  {author} {\bibinfo {author} {\bibfnamefont {T.}~\bibnamefont
  {Lahaye}}, \bibinfo {author} {\bibfnamefont {C.}~\bibnamefont {Menotti}},
  \bibinfo {author} {\bibfnamefont {L.}~\bibnamefont {Santos}}, \bibinfo
  {author} {\bibfnamefont {M.}~\bibnamefont {Lewenstein}}, \ and\ \bibinfo
  {author} {\bibfnamefont {T.}~\bibnamefont {Pfau}},\ }\href {\doibase
  10.1088/0034-4885/72/12/126401} {\bibfield  {journal} {\bibinfo  {journal}
  {Rep. Prog. Phys.}\ }\textbf {\bibinfo {volume} {72}},\ \bibinfo {pages}
  {126401} (\bibinfo {year} {2009})}\BibitemShut {NoStop}%
\bibitem [{\citenamefont {Lahrz}\ \emph {et~al.}(2015)\citenamefont {Lahrz},
  \citenamefont {Lemeshko},\ and\ \citenamefont {Mathey}}]{LarzNJP15}%
  \BibitemOpen
  \bibfield  {author} {\bibinfo {author} {\bibfnamefont {M.}~\bibnamefont
  {Lahrz}}, \bibinfo {author} {\bibfnamefont {M.}~\bibnamefont {Lemeshko}}, \
  and\ \bibinfo {author} {\bibfnamefont {L.}~\bibnamefont {Mathey}},\ }\href
  {\doibase 10.1088/1367-2630/17/4/045005} {\bibfield  {journal} {\bibinfo
  {journal} {New J. Phys.}\ }\textbf {\bibinfo {volume} {17}},\ \bibinfo
  {pages} {045005} (\bibinfo {year} {2015})}\BibitemShut {NoStop}%
\bibitem [{\citenamefont {Lemeshko}(2011)}]{LemeshkoPRA11Optical}%
  \BibitemOpen
  \bibfield  {author} {\bibinfo {author} {\bibfnamefont {M.}~\bibnamefont
  {Lemeshko}},\ }\href {\doibase 10.1103/PhysRevA.83.051402} {\bibfield
  {journal} {\bibinfo  {journal} {Phys. Rev. A}\ }\textbf {\bibinfo {volume}
  {83}},\ \bibinfo {pages} {051402(R)} (\bibinfo {year} {2011})}\BibitemShut
  {NoStop}%
\bibitem [{\citenamefont {Lemeshko}\ and\ \citenamefont
  {Friedrich}(2012)}]{LemFri11OpticalLong}%
  \BibitemOpen
  \bibfield  {author} {\bibinfo {author} {\bibfnamefont {M.}~\bibnamefont
  {Lemeshko}}\ and\ \bibinfo {author} {\bibfnamefont {B.}~\bibnamefont
  {Friedrich}},\ }\href {\doibase 10.1080/00268976.2012.689868} {\bibfield
  {journal} {\bibinfo  {journal} {Mol. Phys.}\ }\textbf {\bibinfo {volume}
  {110}},\ \bibinfo {pages} {1873} (\bibinfo {year} {2012})}\BibitemShut
  {NoStop}%
\bibitem [{\citenamefont {Schmidt}\ \emph {et~al.}(2012)\citenamefont
  {Schmidt}, \citenamefont {Enss}, \citenamefont {Pietil\"{a}},\ and\
  \citenamefont {Demler}}]{Schmidt2011a}%
  \BibitemOpen
  \bibfield  {author} {\bibinfo {author} {\bibfnamefont {R.}~\bibnamefont
  {Schmidt}}, \bibinfo {author} {\bibfnamefont {T.}~\bibnamefont {Enss}},
  \bibinfo {author} {\bibfnamefont {V.}~\bibnamefont {Pietil\"{a}}}, \ and\
  \bibinfo {author} {\bibfnamefont {E.}~\bibnamefont {Demler}},\ }\href
  {http://arxiv.org/abs/1110.1649v2} {\bibfield  {journal} {\bibinfo  {journal}
  {Phys. Rev. A}\ }\textbf {\bibinfo {volume} {85}},\ \bibinfo {pages}
  {021602(R)} (\bibinfo {year} {2012})},\ \Eprint
  {http://arxiv.org/abs/1110.1649} {arXiv:1110.1649} \BibitemShut {NoStop}%
\end{thebibliography}%

\end{document}